\documentclass[]{spie}  

\usepackage[square,numbers,sort&compress,comma]{natbib}
\usepackage{amsmath,amsfonts,amssymb}
\usepackage{graphicx}
\usepackage[export]{adjustbox}
\usepackage{caption}
\graphicspath{{./}{./figures/}}
\usepackage[colorlinks=true, allcolors=blue]{hyperref}
\usepackage{threeparttable}
\usepackage[usenames,dvipsnames,svgnames]{xcolor}

\newcommand{\eq}{\begin{equation*}}
\newcommand{\qe}{\end{equation*}}

\usepackage{aas_macros}
\usepackage[]{natbib}
\usepackage{amsmath}
\usepackage{amssymb}
\usepackage{xspace}
\usepackage{xifthen}
\usepackage{eso-pic}

\newcommand{\reportnum}[2]{
  \AddToShipoutPictureBG*{%
    \AtPageUpperLeft{%
      \hspace{0.65\paperwidth}%
      \raisebox{#1\baselineskip}{%
        \makebox[0pt][l]{\textnormal{#2}}
  }}}%
}

\mathchardef\mhyphen="2D

\newcommand{\roughly}{\ensuremath{ {\sim}\,} }
\newcommand{\gtr}{\ensuremath{ {>}\,} }

\newlength{\dhatheight}

\newcommand{\unit}[1]{\ensuremath{\mathrm{\,#1}}\xspace}

\newcommand{\nm}{\unit{nm}}
\newcommand{\um}{\unit{\mu m}}
\newcommand{\cm}{\unit{cm}}

\newcommand{\second}{\unit{s}}

\newcommand{\e}{\unit{e^{-}}}

\newcommand{\rms}{\unit{rms}}
\newcommand{\pix}{\unit{pix}}

\newcommand{\hz}{\unit{Hz}}
\newcommand{\khz}{\unit{kHz}}
\newcommand{\db}{\unit{dB}}

\newcommand{\rmspix}{\unit{rms/pix}}

\newcommand{\ermspix}{\e \rmspix}

\newcommand{\epixsec}{\unit{\e \pix^{-1} \second^{-1}}}

\newcommand{\erms}{\unit{\e \rms}}
\newcommand{\gain}{\unit{V/V}}
\newcommand{\eadu}{\unit{\e/ADU}}

\newcommand{\texp}{\ensuremath{t_{\rm exp}}\xspace}

\newcommand{\SNR}{\ensuremath{\rm S/N}\xspace}

\title{Design of a Skipper CCD Focal Plane for the SOAR Integral Field Spectrograph}

\reportnum{-2.5}{FERMILAB-CONF-22-484-LDRD-PPD}

\author[1,2,3,*]{Edgar Marrufo Villalpando}
\author[3,2,4]{Alex Drlica-Wagner}
\author[5]{Marco Bonati}
\author[3]{Abhishek Bakshi}
\author[6]{Vanessa Bawden de Paula Macanhan}
\author[5]{Braulio Cancino}
\author[3]{Gregory E.\ Derylo}
\author[3]{Juan Estrada}
\author[3]{Guillermo Fernandez Moroni}
\author[6]{Luciano Fraga} 
\author[7]{Stephen Holland}
\author[3]{Michelle J.\ Jonas} 
\author[9,3]{Agustín Lapi} 
\author[5]{Peter Moore}
\author[8]{Andrés A.\ Plazas Malagón}
\author[3]{Leandro Stefanazzi}
\author[3]{Javier Tiffenberg}

\affil[1]{\small Department of Physics, University of Chicago, Chicago, IL 60637, USA}
\affil[2]{\small Kavli Institute of Cosmological Physics, University of Chicago, Chicago, IL 60637, USA}
\affil[3]{\small Fermi National Accelerator Laboratory, Batavia, IL 60510, USA}
\affil[4]{\small Department of Astronomy \& Astrophysics, University of Chicago, Chicago, IL 60637, USA}
\affil[5]{\small Cerro Tololo Inter-American Observatory, NSF’s National Optical-Infrared Astronomy Research Laboratory, Casilla 603, La Serena, Chile}
\affil[6]{\small Laborat\'orio Nacional de Astrof\'isica LNA/MCTI, 37504-364, Itajub\'a, MG, Brazil}
\affil[7]{\small Lawrence Berkeley National Laboratory, One Cyclotron Rd, Berkeley, CA 94720, USA}
\affil[8]{\small Department of Astrophysical Sciences, Princeton University, Princeton, NJ 08544, USA}
\affil[9]{\small Departamento de Ingenier\'ia El\'ectronica y de Computadoras, Universidad Nacional del Sur, Bahia Blanca, Argentina}

\authorinfo{$^*$~emarrufo@uchicago.edu}

\begin{document} 
\maketitle

\begin{abstract}
We present the development of a Skipper Charge-Coupled Device (CCD) focal plane prototype for the SOAR Telescope Integral Field Spectrograph (SIFS). This mosaic focal plane consists of four 6k $\times$ 1k, 15 $\mu$m pixel Skipper CCDs mounted inside a vacuum dewar. We describe the process of packaging the CCDs so  that they can be easily tested, transported, and installed in a mosaic focal plane. We characterize the performance of $\sim 650 \mu$m thick, fully-depleted engineering-grade Skipper CCDs in preparation for performing similar characterization tests on science-grade Skipper CCDs which will be thinned to 250$\mu$m and backside processed with an antireflective coating. We achieve a single-sample readout noise of $4.5 \ermspix$ for the best performing amplifiers and sub-electron resolution (photon counting capabilities) with readout noise $\sigma \sim 0.16 \ermspix$ from 800 measurements of the charge in each pixel. We describe the design and construction of the Skipper CCD focal plane and provide details about the synchronized readout electronics system that will be implemented to simultaneously read 16 amplifiers from the four Skipper CCDs (4-amplifiers per detector). 
Finally, we outline future plans for laboratory testing, installation, commissioning, and science verification of our Skipper CCD focal plane.
\end{abstract}


\section{INTRODUCTION}
\label{sec:intro}  
The Southern Astrophysical Research (SOAR) Telescope Integral Field Spectrograph (SIFS) is an astronomical spectrograph equipped with an optical fiber-lenslet integral field unit (IFU) \cite{10.1117/12.857698}. SIFS was developed and constructed in Brazil by Laboratório Nacional de Astrofísica, Ministry of Science, Technology and Innovation (LNA/MCTI) in partnership with Instituto de Astronomia, Geofísica e Ciências Atmosféricas, Universidade de São Paulo (IAG/USP) for the 4.1-m SOAR telescope, located on Cerro Pachón in the Coquimbo Region of Chile. Integral field spectroscopy provides a spectrum for each spatial element in a two dimensional field; in the case of SIFS, data products consists of three-dimensional data cubes with axes of right accession, declination, and wavelength. The main science driver that has guided the design of SIFS is the study of complex extended objects such as \textsc{H\,ii} clouds and star-forming regions in galaxies by providing simultaneous spectra of such extended objects. The high spatial resolution of SIFS is of particular importance for the study of velocity fields and ionization structure in \textsc{H\,ii} regions, active galactic nuclei (AGN) \cite{da_Silva_2020}, and planetary nebulae \cite{10.1117/12.461977,10.1117/12.857698}.  

Astronomical spectrographs disperse light over a large detector area. For faint astronomical sources, this results in low signal-to-noise in each detector pixel. 
This is a particularly common case in observations with SIFS, where signal-to-noise often is a limiting factor in scientific measurements (e.g., when constructing a velocity map far from the nucleus of an AGN) \cite{10.1117/12.857698,da_Silva_2020}.
In the low-signal-to-noise regime, detector readout noise can be an important contribution to the overall noise in an observation, affecting the sensitivity of spectroscopic measurements. Skipper CCDs offer a novel solution to the problem of detector readout noise by using a floating gate output stage to perform repeated measurements of the charge in each pixel. These measurements can be combined to reduce readout noise relative to a single measurement and achieve sub-electron resolution, i.e., single electron/photon counting. The Skipper CCD concept as a photosensitive detector was proposed in 1990 \cite{Janesick:1990,10.1117/12.19457}; however, in early demonstrations of this technology, the readout noise deviated from the theoretical expectation at $\sim 0.5 \ermspix$, which prevented single photon counting \citep{Janesick:1990}. In contrast, modern Skipper CCDs have achieved an order of magnitude lower readout noise and stable performance over a large area detector \cite{Tiffenberg:2017}.  

Here we present the development of a Skipper CCD focal plane prototype for SIFS \cite{10.1117/12.461977}. Our goal is to demonstrate the low-readout noise capabilities of modern Skipper CCDs (targeting a $\roughly7\times$ reduction in the readout noise of SIFS), while exposing Skipper CCDs to the full complexities of astronomical spectroscopy for the first time. The Skipper CCD focal plane will replace the current SIFS dewar and detector with a dewar  housing a mosaic focal plane of four 6k $\times$ 1k Skipper CCDs. We will also replace the readout electronics with low-threshold readout boards designed for Skipper CCDs (Section \ref{sec:electronics}). The reduction in readout noise that Skipper CCDs offer will improve SIFS' signal-to-noise when imaging faint sources in the low-signal, low sky-background regime such as N\,{\sc ii} or S\,{\sc ii} spectral lines. In this paper, we describe the process for designing and testing the Skipper CCD focal plane for SIFS.

\section{THE SKIPPER CCD for Astronomy}

 Skipper CCDs use a floating gate output stage to allow for multiple, non-destructive measurements of the charge in each pixel. Tiffenberg et al.\ (2017) \cite{Tiffenberg:2017} demonstrated that Skipper CCDs designed at Lawrence Berkeley National Laboratory (LBNL), fabricated at Teledyne DALSA, and packaged at Fermi National Accelerator Laboratory (Fermilab) could implement a floating-gate output stage, a small capacitance sense node, and isolation from parasitic noise sources to perform repeated, nondestructive measurements of the charge in each pixel of a thick, fully depleted Skipper CCD, achieving a readout noise of $0.068 \ermspix$ \citep{Tiffenberg:2017}. Subsequent testing has demonstrated that this new generation of Skipper CCDs can achieve readout noise values of  $< 0.039 \ermspix$ \cite{Cancelo:2020}. The readout noise achieved after $N_{\rm samp}$ independent measurements, $\sigma_N$, follows the expected behavior from Gaussian statistics: 
 
 \begin{equation}
    \sigma_N = \frac{\sigma_{1}}{\sqrt{N_{\rm samp}}}
    \label{eqn:noise}
\end{equation}

\noindent where $\sigma_1$ is the single-sample readout noise (the standard deviation of pixel values with a single charge measurement per pixel) and $N_{\rm samp}$ is the number of measurements performed for each pixel \cite{Cancelo:2020}.

\begin{table}[t]
\centering
\caption{\label{tab:skipper}
Summary of Skipper CCD Detector Performance from the Literature
}
\begin{tabular}{l c c c c}
\hline
Characteristic  & Value  & Unit & Reference\\
\hline \hline
Single-Sample Readout Noise & 2.5 & \ermspix & \cite{10.1117/1.JATIS.7.1.015001} \\
Multi-Sample Readout Noise & 0.039 & \ermspix & \cite{10.1117/1.JATIS.7.1.015001} \\
Dark Current & $6.82 \times 10^{-9}$ & \epixsec & \cite{Barak_2022} \\
Spurious Clock-Induced Charge & $1.52 \times 10^{-4}$ & \e /pix/frame & \cite{Barak_2022} \\
Quantum Efficiency ($<900\nm$) & $>75\%$ & ... & \cite{10.1117/12.2562403} \\
Charge Transfer Inefficiency & $1.9 \times 10^{-6}$ & ... & \cite{10.1117/12.2562403} \\
Full-Well Capacity & $>33,000$  & \e & \cite{10.1117/12.2562403} \\
\hline
\end{tabular}
\end{table}

In addition to having the capability to dynamically tune the readout noise on a pixel-by-pixel basis, Skipper CCDs maintained the well-characterized performance of the thick, fully depleted, $p$-channel CCDs developed by LBNL for astronomy (i.e., stability, linear response, large dynamic range, high QE in the red and near infra-red regimes, and radiation tolerance) \cite{10.1117/12.2562403,Dawson:2008}. Skipper CCDs have been extensively characterized in the literature (see Table \ref{tab:skipper}), demonstrating unprecedented performance for particle physics and astronomical applications. In particular, we note the first optical characterizations of the Skipper CCD for cosmological applications \cite{10.1117/12.2562403}, which informed the plans to test the Skipper CCD in a realistic astronomical observing scenario. In \cite{10.1117/12.2562403}, we demonstrated that a backside illuminated Skipper CCD could achieve relative QE $>75\%$ from 450\nm to 900\nm, with relative QE $\gtrsim 30\%$ out to 1\um. 
When compared to the absolute QE of LBNL CCDs developed for the Dark Energy Camera (DECam) and the Dark Energy Spectroscopic Instrument (DESI), we expect that it is possible to achieve QE $>80\%$ from 450\nm to 950\nm. Furthermore, we showed that the Skipper CCD achieved a full-well capacity of $33,000$ $\e$ with a shallow voltage configuration, and  we expect that by optimizing clock voltages, the Skipper CCD can achieve values $\gtr150,000$ $\e$, based on the performance of other thick, high resistivity CCDs from LBNL \cite{Flaugher:2015}.

\section{SOAR Integral Field Spectrograph}
\label{sec:SIFS}

\subsection{System Characteristics}
The SOAR Telescope is a 4.1-m Ritchey-Chretien alt-azimuth telescope that  provides high-precision pointing with a large instrument payload capacity and excellent angular resolution with tip-tilt correction \cite{10.1117/12.857698,10.1117/12.856593}. SIFS is a fiber-fed integral field spectrograph consisting of 1300 fibers covering a 15 $\times$ 7.8 arsec$^{2}$ field-of view with an angular resolution of 0.30 arsec/fiber. This allows SIFS to perform spatially resolved 2D spectroscopy of complex extended objects and crowded fields. SIFS has wide wavelength coverage, from 350\,nm to 1000\,nm, while allowing for simultaneous spectra to be taken of all parts of moderately extended objects such as H\,{\sc ii} regions. Furthermore, SIFS uses volume phase holographic (VPH) transmission gratings that offer a range of spectral resolution ($R = \delta \lambda / \lambda$) from $5000 < R < 30000$, which is sufficient to measure stellar metallicities at $R > 6000$ and abundance ratios, such as H$ \alpha$/Fe, at $R > 15000$  \cite{10.1117/12.461977,10.1117/12.857698,10.1117/12.856593}. For example,  Palumbo et al.\ (2020) used SIFS to measure redshifts, velocity fields, which were constructed by sampling object spectra at many discrete points, specific start formation rates, and surface mass densities of compact dwarf starburst galaxies in the Local Volume \cite{Palumbo:2020}. Patr\'icia da Silva at al.\ (2020) presented a multi-wavelength study of the nuclear region of NGC 613, which contains an AGN. In this study, SIFS is used to obtain a data cube of the central region of NGC 613 with a spatial resolution of 0.3 arsec $\times$ 0.3 arsec; the data is used to study H\,{\sc ii} star-forming regions in the circumnuclear ring of the galaxy \cite{da_Silva_2020}.

\begin{figure}[t!]
    \centering
    \includegraphics[width=0.95\textwidth]{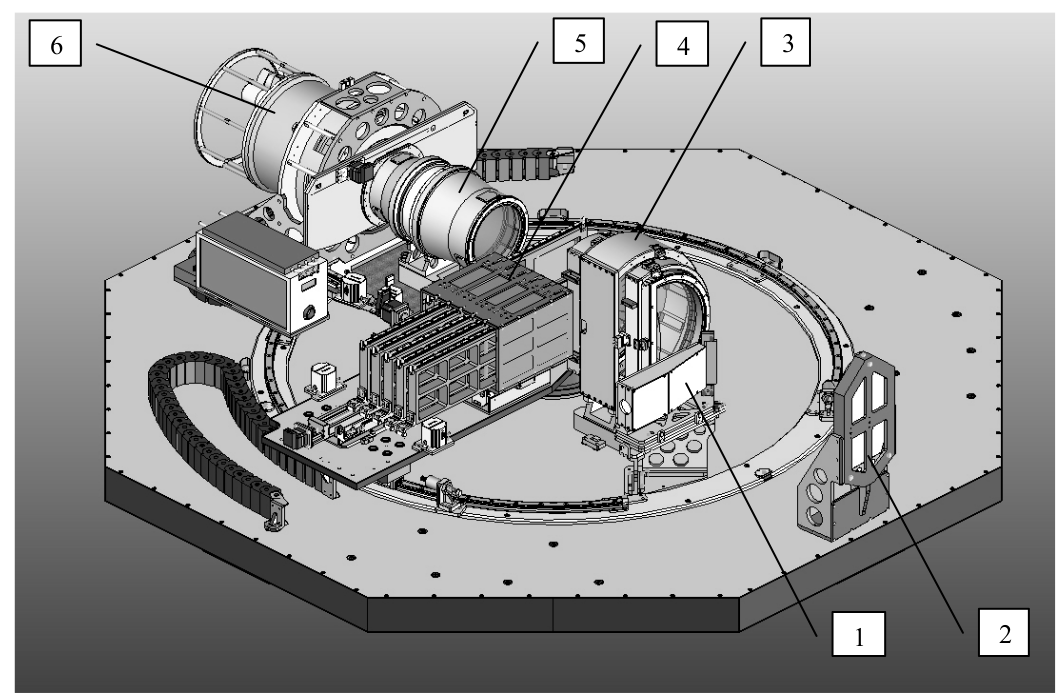}
    \vspace{1em}
    \caption{SIFS bench top spectrograph. (1) Output unit support and fold mirror. (2) Collimator mirror. (3) Collimator lenses. (4) VPH grating mechanism. (5) Camera. (6) Cryostat containing detector. (Diagram taken from \cite{10.1117/12.856593}). }
    \label{fig:spectrograph}
\end{figure}

SIFS is composed of three main subsystems: the fore-optics, the IFU, and the bench spectrograph. The fore-optics consist of a magnification lens group followed by a field lens, which magnifies the focal plane of the telescope to the scale that is required by the micro lens array \cite{10.1117/12.857698}. 
SIFS employs a lenslet, fiber IFU system which consists of 1300 lenslets and optical fibers arranged into a $50 \times 26$  array and encapsulated in 26 furcation tubes. The optical fibers (Polymicro ``blue" fiber) have a spectral transmission $\gtr 70\%$ for wavelengths between 350\,nm and 950\,nm \cite{10.1117/12.857698,10.1117/12.856593}. The bench spectrograph (Fig. \ref{fig:spectrograph}) is installed on the platform of the SOAR telescope, supported by six sets of pneumatic supports for vibration isolation. The bench spectrograph includes the output unit support, the collimator lens group and a spherical mirror, the VPH gratings exchanging mechanism, the camera and detector \cite{10.1117/12.856593}. The SIFS Skipper CCD focal plane will house four Skipper CCDs and be mounted in a replacement cryostat  that matches the mechanical dimensions and focal distance of the original (item 6 in Fig.~\ref{fig:spectrograph}). The modular design of the cryostat simplifies the time, cost and risk of installation since no mechanical or optical modifications are required; furthermore, there is not need to replace any other components of the bench spectrograph.

\subsection{Expected Improvements from a Skipper CCD Focal Plane}

\begin{table}[t]
\centering
\caption{\label{tab:sifs_detector_performance}
SIFS Reference Parameters
}
\begin{tabular}{l c c c }
\hline
Characteristic  & Value  & Unit\\
\hline \hline
Signal Rate & 0.01191 & \epixsec \\
Background Rate & 0.0079 & \epixsec  \\
Dark Current & $3.2 \times 10^{-3}$ & \epixsec  \\
Readout Noise & 5.2 & \ermspix\\
Npix & $3.2 \times 2.1$ & pix  \\
Readout Time & 22 & secs \\
\hline
\end{tabular}
\end{table}

The SIFS bench spectrograph currently uses a thinned CCD231-84 from Teledyne e2v that has a format of 4096 $\times$ 4112 pixels and a pixel size of $15\, \mu$m. The detector is read through two channels in 22s with a readout noise of $5.2 \ermspix$ and a gain of $2.0 \e$/ADU. Table \ref{tab:sifs_detector_performance} shows some characteristic parameters that influence the SIFS performance, including the SIFS detector performance. We note that an evident improvement from the Skipper CCD focal plane will be the single sample per pixel readout noise. Assuming the Skipper CCD can reach a single sample readout noise of 3.5 \ermspix, based on the performance of similar Skipper CCDs  \cite{10.1117/12.2562403,Tiffenberg:2017, Rodrigues:2020}, there will be a $ \sim 40\%$ improvement in readout noise when comparing with SIFS's current detector. 

\begin{figure}[t!]
    \centering
    \includegraphics[width=0.75\textwidth]{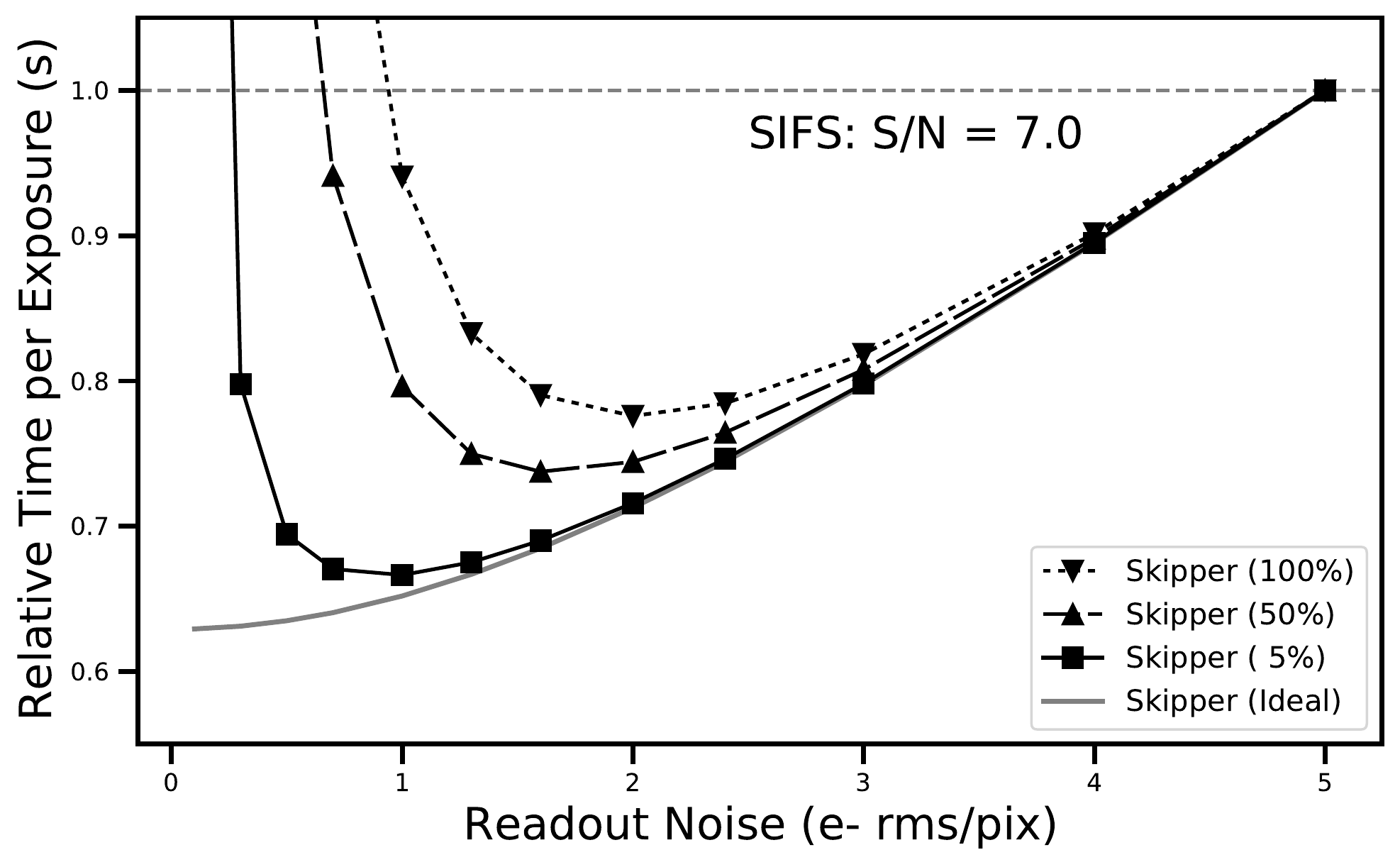}
    \caption{Relative exposure times as a function of detector readout noise at a fixed S/N. Curves represent 100$\%$, 50$\%$, and 5$\%$ of the detector being read out.}
    \label{fig:fixed_SN}
\end{figure}

In Fig.~\ref{fig:fixed_SN}, we explore the observation time required to reach a fixed signal-to-noise (S/N) ratio as a function of readout noise for SIFS. First, we consider the S/N ratio for an astronomical observation, which can be interpreted as the sensitivity to a given source and is defined as the ratio between the number of counts contributed by the source and the total noise in the observation. Mathematically, S/N is expressed as 

\begin{align}
\SNR = \frac{R_{\rm src} \texp}{\Sigma_{\rm tot}}
\label{eqn:snr}
\end{align}

{\setlength{\parindent}{0cm}
where $\Sigma_{\rm tot}$ represents the total noise in the observation, which includes the total electron rates (\epixsec) from the source, background, and dark contributions integrated over pixel elements ($R_{\rm src}, R_{\rm bkg}, R_{\rm dark}$), added in quadrature, assuming uncorrelated pixel measurements, i.e. $\Sigma_{\rm tot} = {\sqrt{ (R_{\rm src}+R_{\rm bkg} + R_{\rm dark})\texp + N_{\rm samp} \sigma_{ N}^2}}$. Here, $N_{\rm samp} \sigma_{N}^{2}$ represents the readout noise of a Skipper CCD, which is an intrinsic property of the detector, independent of the integration time, and tunable or scalable by the number of samples per pixel, $N_{\rm samp}$. The readout noise is quantified in units of the root-mean-square (rms) fluctuations in the number of electrons per pixel. Fig.~\ref{fig:fixed_SN}, constructed with the values from Table \ref{tab:sifs_detector_performance}, shows the expected improvement in the reduction of observation time (exposure + readout time) to reach a target signal-to-noise for different percentages of the detector being read out, i.e., region of interest functionality (Section \ref{sec:outlook}). 
\smallbreak

}

\section{SKIPPER CCD DETECTORS}
\label{sec:detectors}

\subsection{Detector architecture and characteristics}
\begin{figure}[t!]
  \begin{center}
    \includegraphics[width=0.4\textwidth]{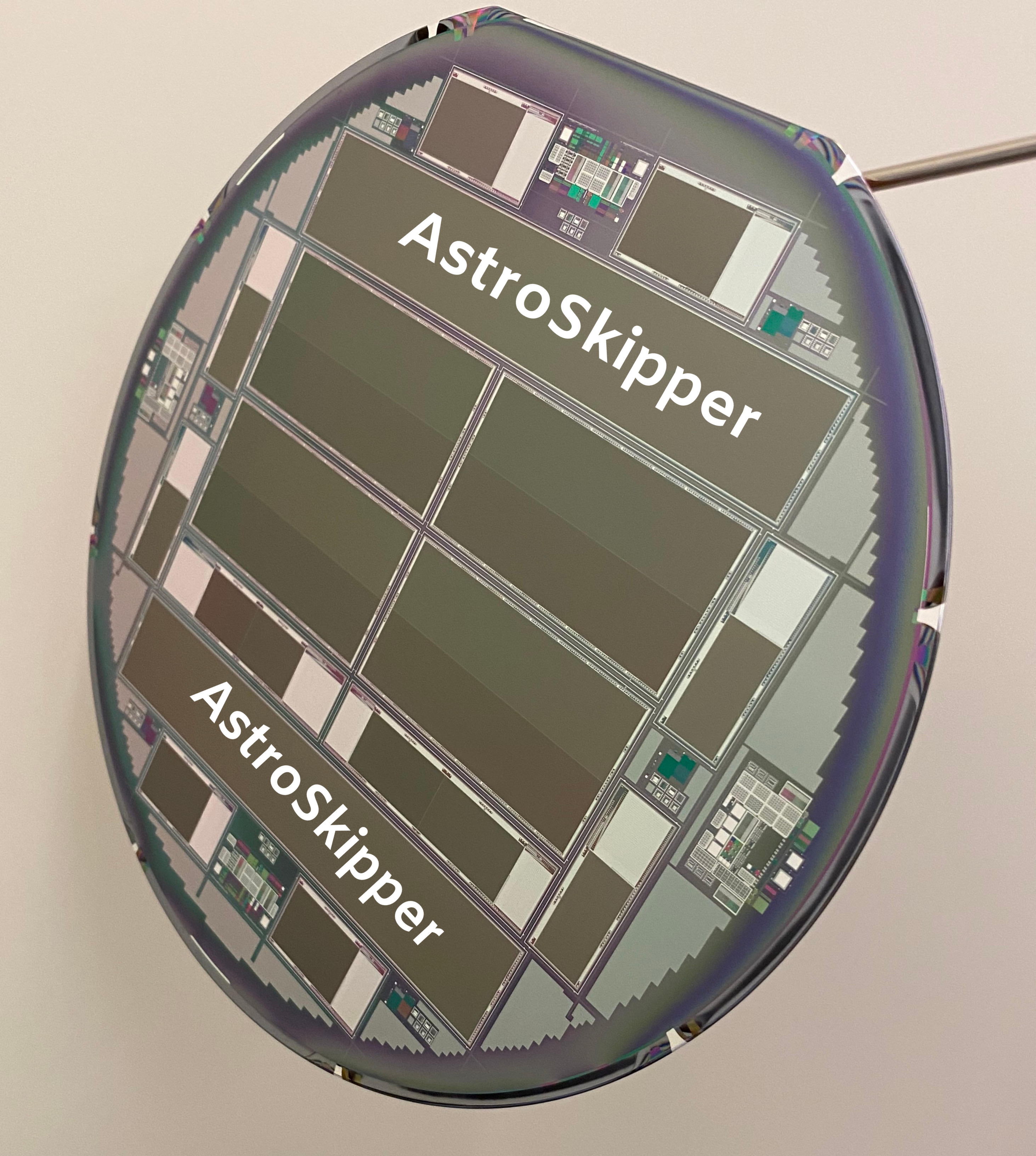}
    \includegraphics[width=0.58\textwidth]{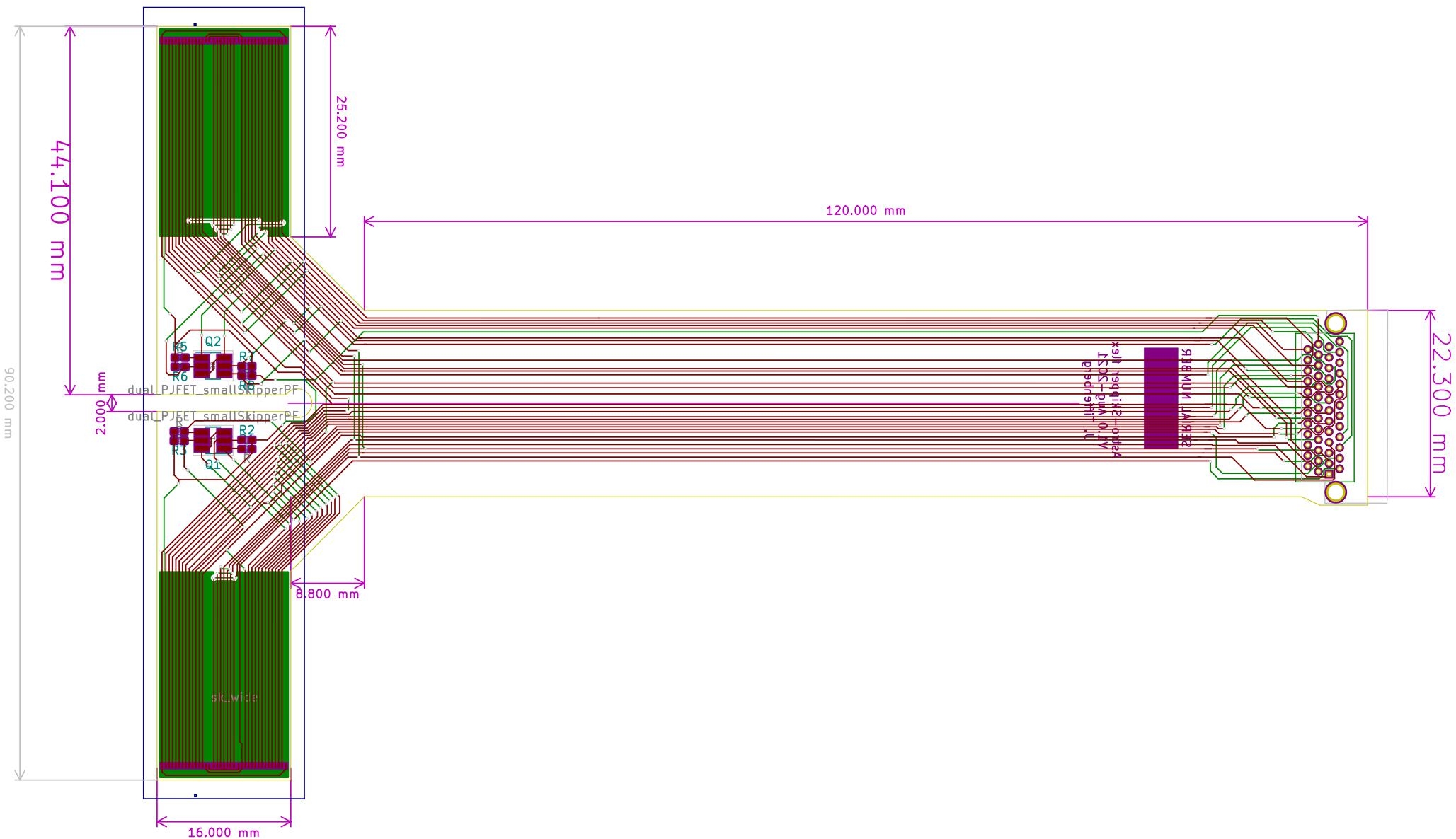}
  \end{center}
  \caption{\label{fig:wafer}
Left: Silicon wafer containing 16 Skipper CCDs for different Fermilab R\&D projects. We will be using the 6k $\times$ 1k pixel CCDs with four amplifiers for the Skipper CCD focal plane (labeled ``AstroSkipper''). Right: AstroSkipper flex cable with active components, i.e., two JFETs and resistors. The flex cable uses an Omnetics connector (right edge of the cable). 
}
\end{figure}

The four backside illuminated Skipper CCDs that will be used for the SIFS Skipper CCD focal plane come from a fabrication run for Fermilab R\&D projects. The Skipper CCDs are $p$-channel detectors fabricated on high resistivity ($\sim$5~k$\Omega\cm$), $n$-type silicon. The wafers are fabricated at a thickness of $\sim 650 \um$, and will be thinned, and backside processed with an antireflective coating at the LBNL Microsystems Laboratory. Each silicon wafer contains 16 Skipper CCDs (Fig. \ref{fig:wafer}) with different readout and size configurations. For the SIFS Skipper CCD focal plane, we will use standard wide-format Skipper CCDs (6k $\times$ 1k, 15 $\mu$m pixels) with four amplifiers (``AstroSkipper'' in Fig. \ref{fig:wafer}). The pixel scale of these detectors matches that of the existing SIFS e2v detector, and a mosaic of four 6k $\times$ 1k Skipper CCD detectors will be used to cover the full $\sim$4k $\times$ 4k pixel area of the current SIFS detector.




\subsection{Detector Packaging}
The AstroSkipper CCD detector package has two main components: a flexible cable for carrying electrical signals to/from the CCD and a mechanical foot for mounting the CCD to the focal plane. The flexible cable carries the bias voltages, clock voltages, and video signals and it has two junction-gate field-effect transistors (JFETs), four 20 k$\Omega$ resistors, and a 51-pin Omnetics connector. We use two high performance dual LSJ689-SOT-23, $p$-channel JFETs that provide ultra-low noise ($\sim 2.0$nV/$\sqrt{\rm Hz}$ ). 

\begin{figure}[t!]
  \begin{center}
    \includegraphics[width=0.24\textwidth]{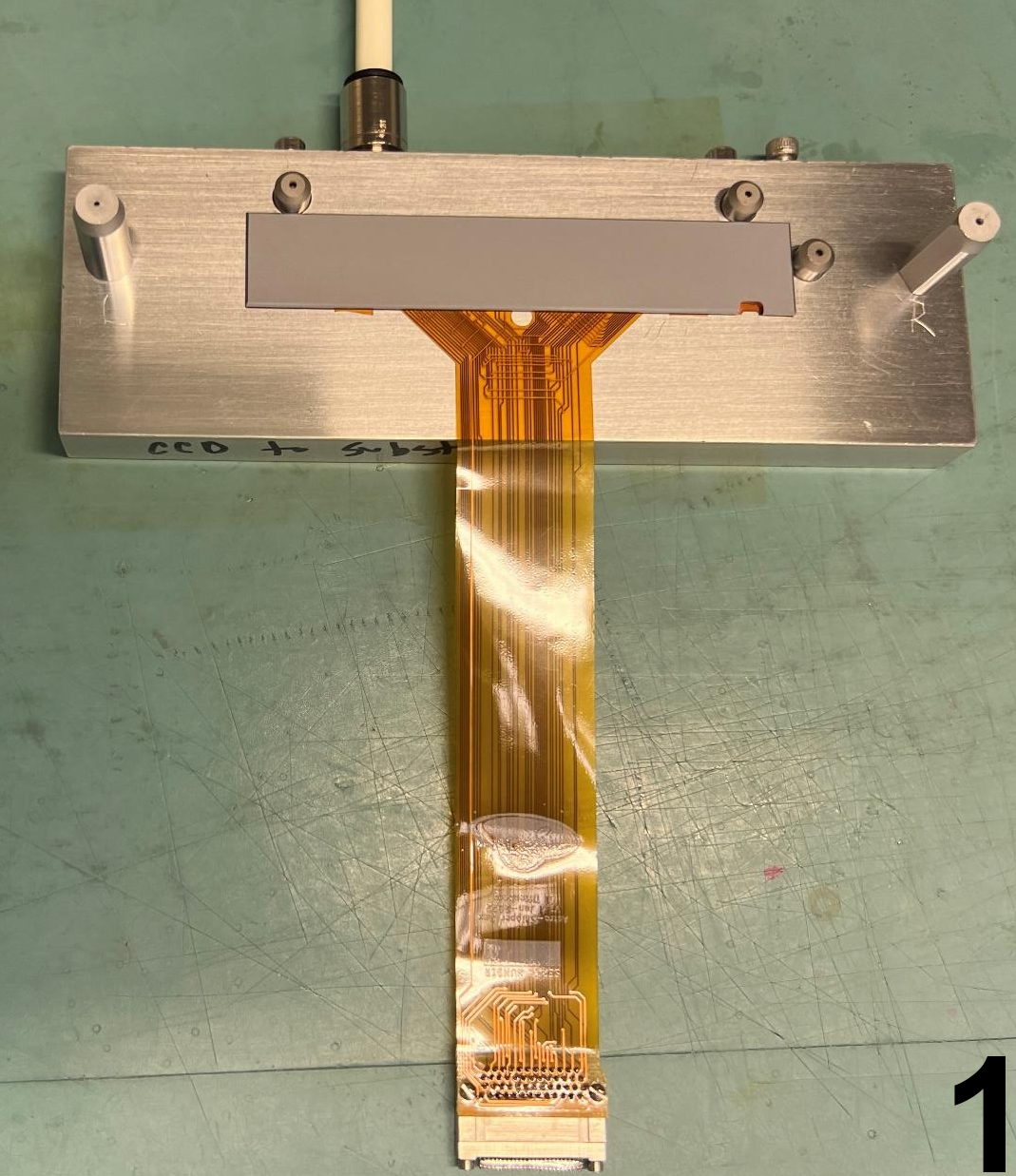}
    \includegraphics[width=0.325\textwidth]{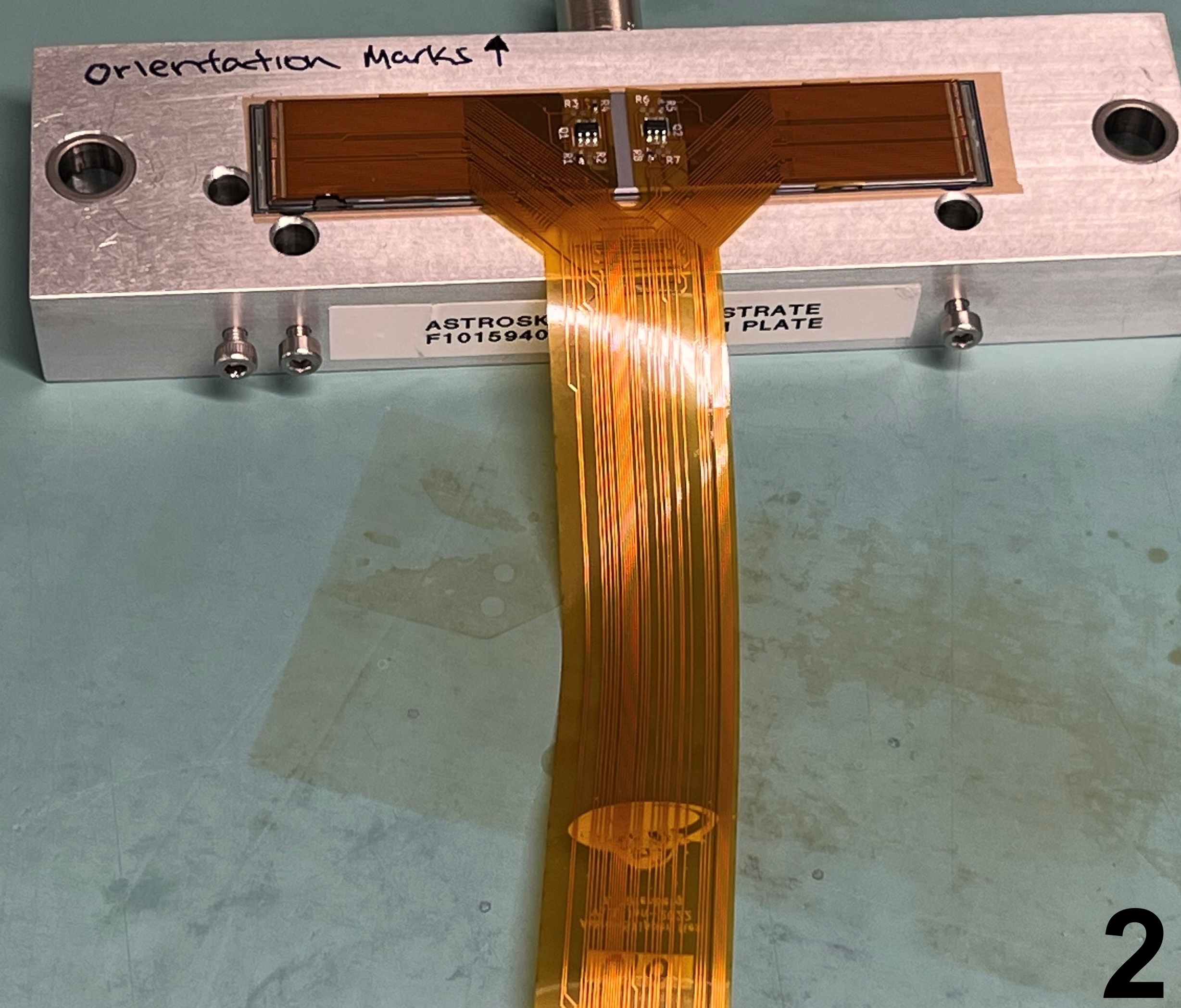}
     \includegraphics[width=0.285\textwidth]{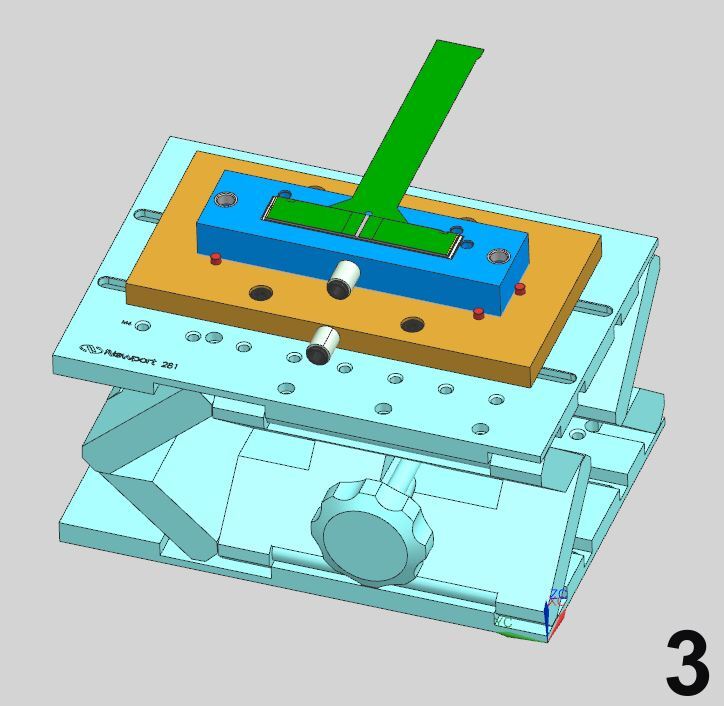}
      \includegraphics[width=0.342\textwidth]{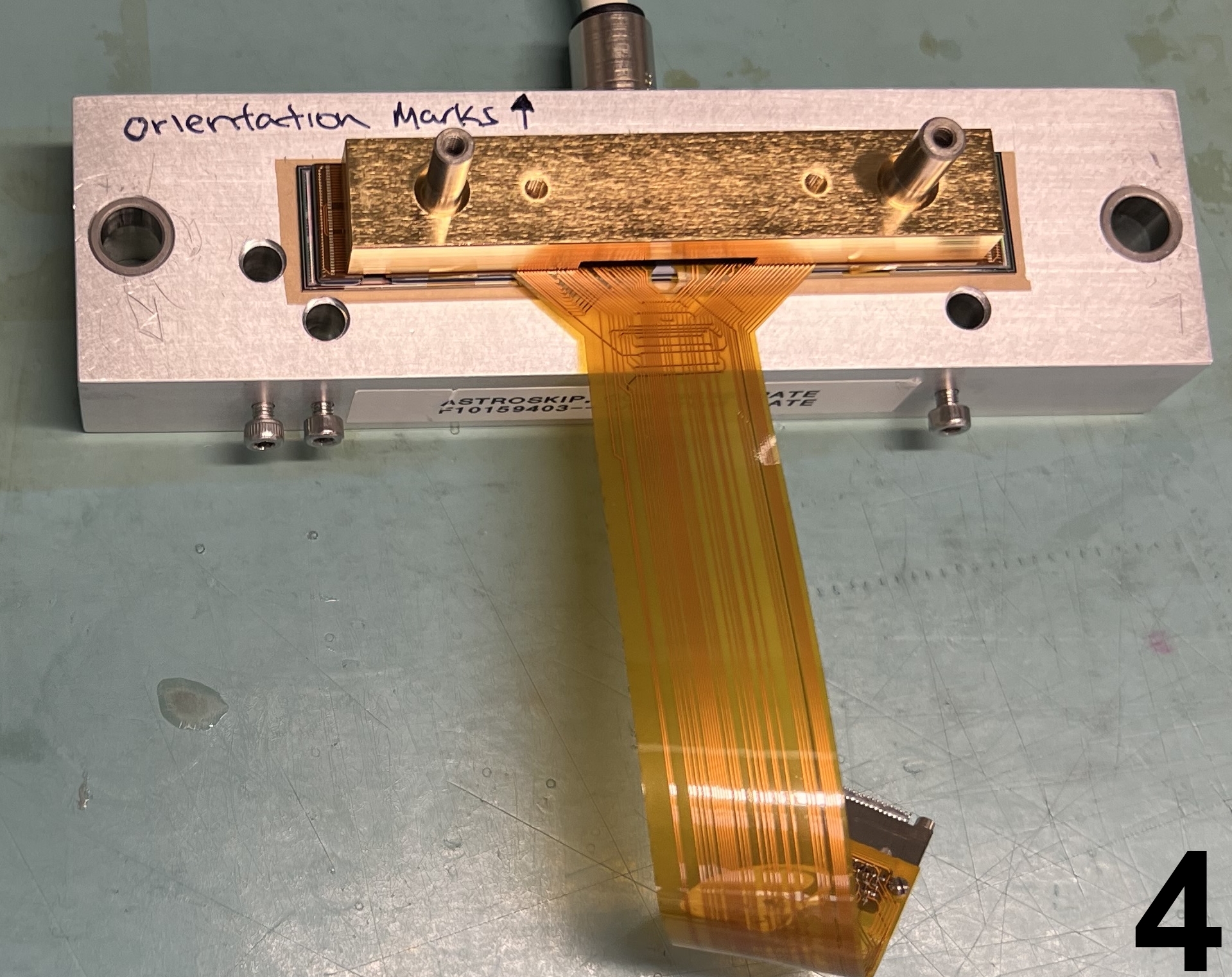}
       \includegraphics[width=0.37\textwidth]{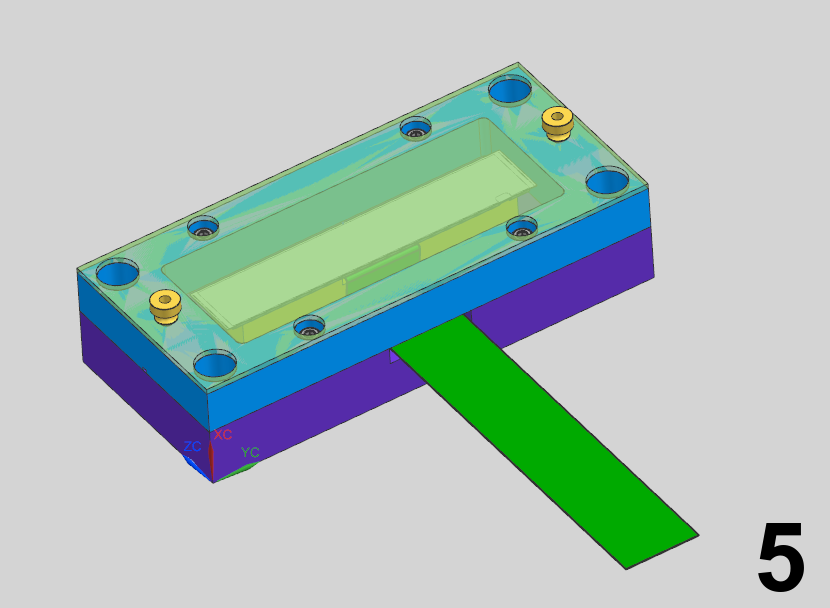}
  \end{center}
  \caption{\label{fig:packaging} Skipper CCD packaging process. \textbf{(1)} Flex cable, with Omnetics connector at the end, is glued to the Si substrate. \textbf{(2)} The CCD, placed on the blue fixture, is attached to the flex-Si substrate assembly. \textbf{(3)} Wirebonding between CCD pads and flex cable. \textbf{(4)} CCD foot, gold-plated piece, is glued to the CCD. \textbf{(5)} CCD package is installed into a carrying box with removable lid for storage, testing, and transport. 
}
\end{figure}

The mechanical packaging process (Fig. \ref{fig:packaging}) starts by (1) attaching the flex cable to a  90.74mm $\times$ 16.36mm $\times$ 1.00mm silicon (Si) substrate piece. A mechanical fixture is used to align the Si substrate and a vacuum system that runs through the fixture is used to apply suction and hold the substrate in place. The flex cable is attached to the Si substrate using two pyralux sheets. The Si substrate--pyralux--flex cable assembly is cured at $191^{\circ}$\,C for 3.5\,hours under vacuum. (2) The CCD is attached to the Si substrate--flex cable assembly. Another fixture is used to align the CCD with the the Si substrate, and double-sided tape is used to secure the CCD to the Si substrate while epoxy is applied \cite{Derylo:2006}. Epotek 301-2 epoxy is dispensed through a small cutout in the edge of the Si substrate near one of the corners and is allowed to fill the region between the CCD and the Si substrate. The epoxy is allowed to cure for 48\,hours. (3) Wirebonding is performed between the pads on the CCD and flex cable.  (4) We attach the CCD assembly to a gold-plated invar foot that serves as the structure for mounting the packaged Skipper CCDs on the focal plane. The fixture holding the CCD  is placed onto the foot gluing plate; Masterbond EP21TCHT-1 epoxy is used to attach the CCD assembly and the foot and is allowed 18 to 24\,hours to cure.  (5) The packaged AstroSkipper is placed within an aluminium carrying box for storage and transport. The carrying box has a removable cover and is compatible with our CCD testing stations.  The storage box can be directly mounted to the cold-finger plate inside the testing vacuum chambers and the storage box cover can be removed for CCD testing.  To date, we have successfully packaged and tested three thick AstroSkipper CCDs at Fermilab.

\begin{figure}[t!]
    \centering
    \includegraphics[width=0.45\textwidth]{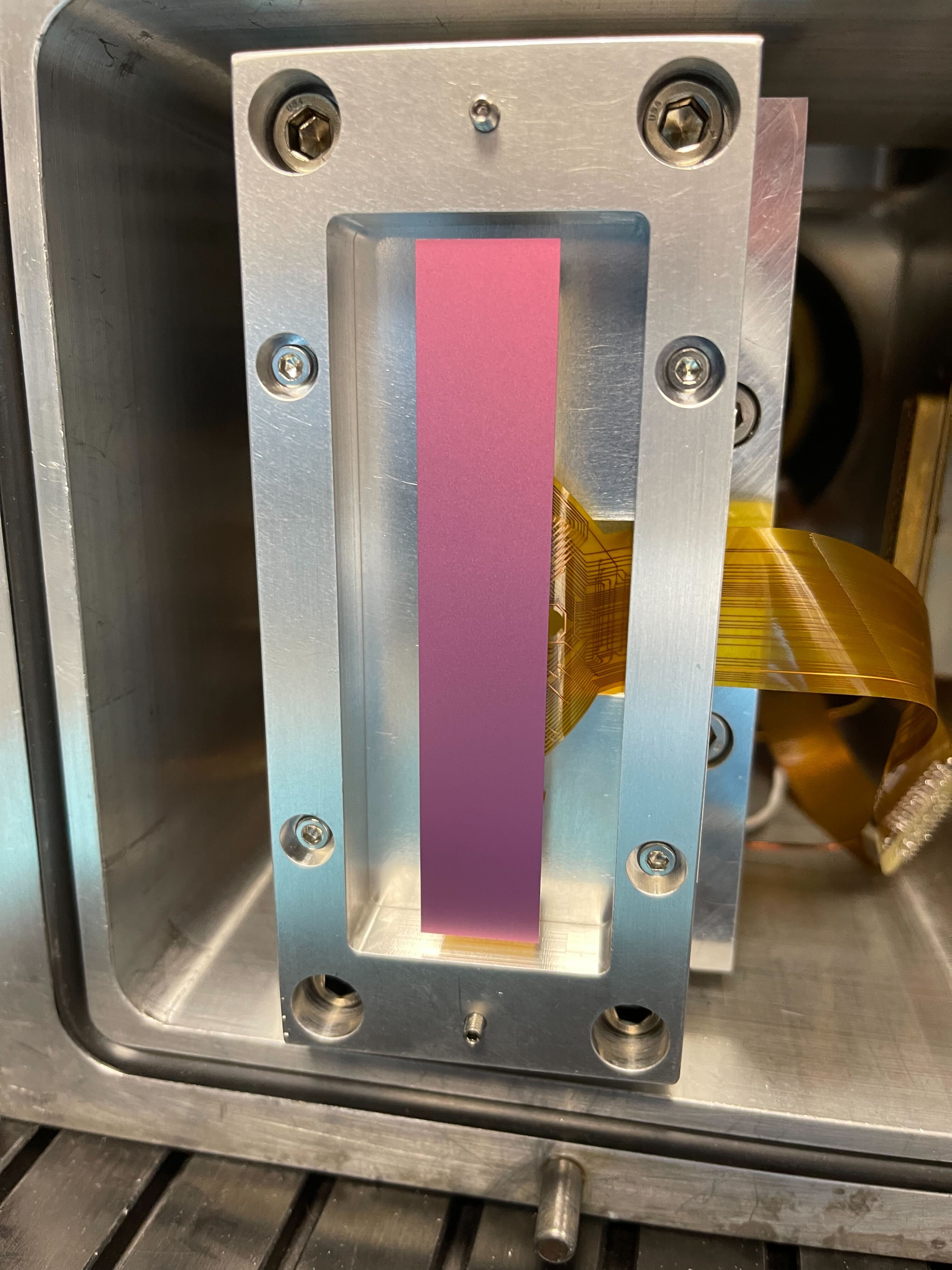}
    \vspace{1em}
    \caption{Thick AstroSkipper CCD packaged at Fermilab (detector is inside carrying box). The AstroSkipper CCD and carrying box have been mounted inside of the vacuum chamber for testing.}
    \label{fig:detector}
\end{figure}

\subsection{Testing Stations}

We have assembled two similar testing stations for characterizing the performance of the packaged AstroSkipper CCDs. The detectors are installed in closed-cycle vacuum dewars. One of the dewars has a fused silica window that can be used to illuminate the detector for optical characterization tests. The AstroSkipper carrying box attaches to an aluminium plate that is screwed to a copper cold finger. The systems are cooled by a closed-system cryogenic cooler; the temperature in our systems is maintained stable at 140K by a Laskeshore temperature controller. The readout chain consists of a second-stage flex cable, an internal dewar board (IDB), a vacuum interference board (VIB), an output dewar board (ODB), and a low-threshold acquisition (LTA) board (Section \ref{sec:electronics}). The IDB allows for selecting 4 out of 16 channels for readout. Since we are testing the performance of individual detectors (each detector has four amplifiers), we select the desired amplifiers with the IDB and use a single LTA to read out a single detector. We are currently designing a second stage flex cable for Fermilab testing that will select the four channels, eliminating the IDB from the readout chain and potentially eliminating noise sources.


\subsection{Detector Performance}
We implement the following tests to characterize the performance of the packaged AstroSkipper CCDs: measuring the readout noise, which involves quantifying electron peaks and verifying that the readout noise behavior follows Eq. \eqref{eqn:noise}, and measuring the charge transfer inefficiency (CTI).

\begin{figure}[t!]
    \centering
    \includegraphics[width=0.47\textwidth]{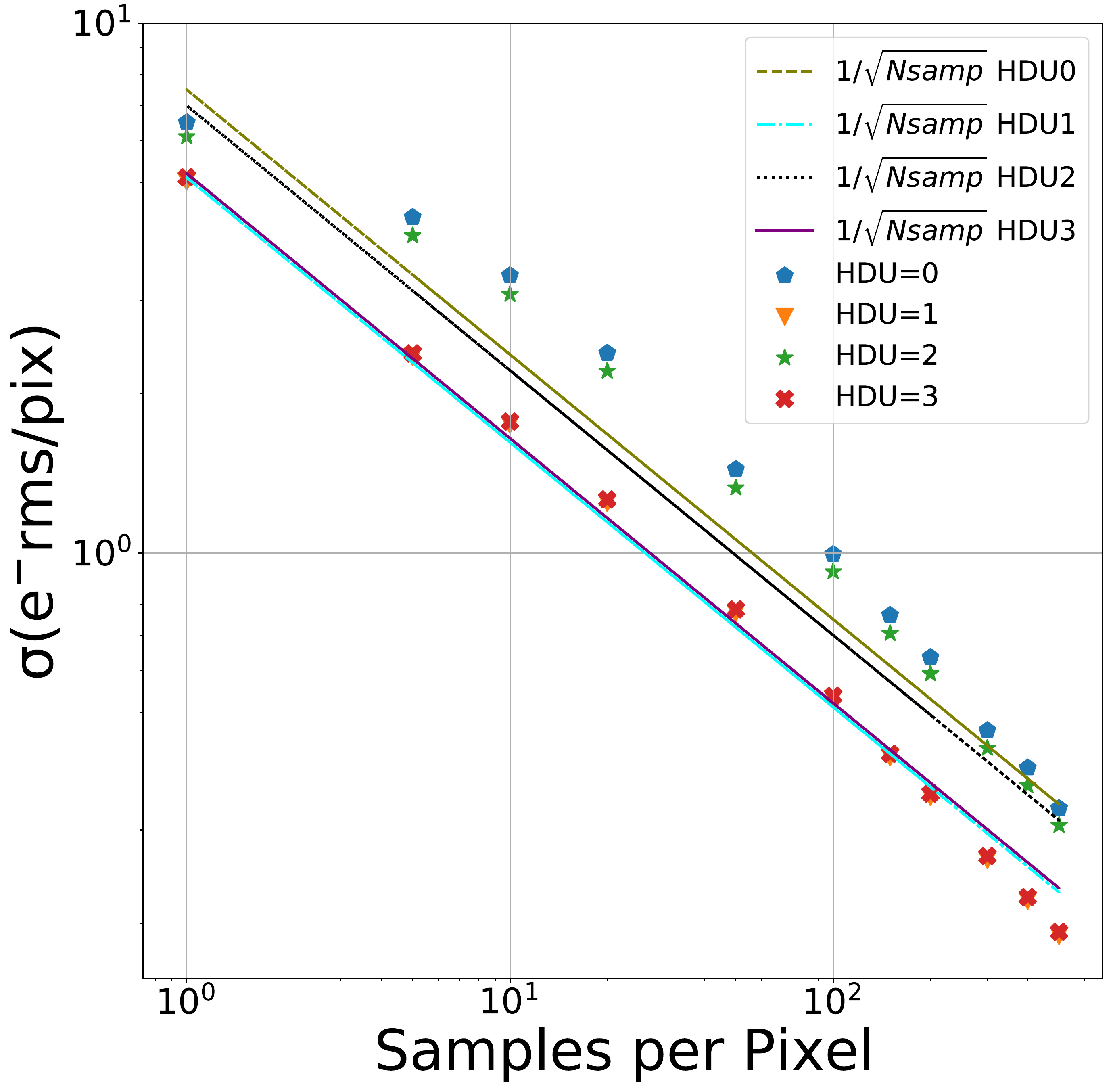}
    \includegraphics[width=0.49\textwidth]{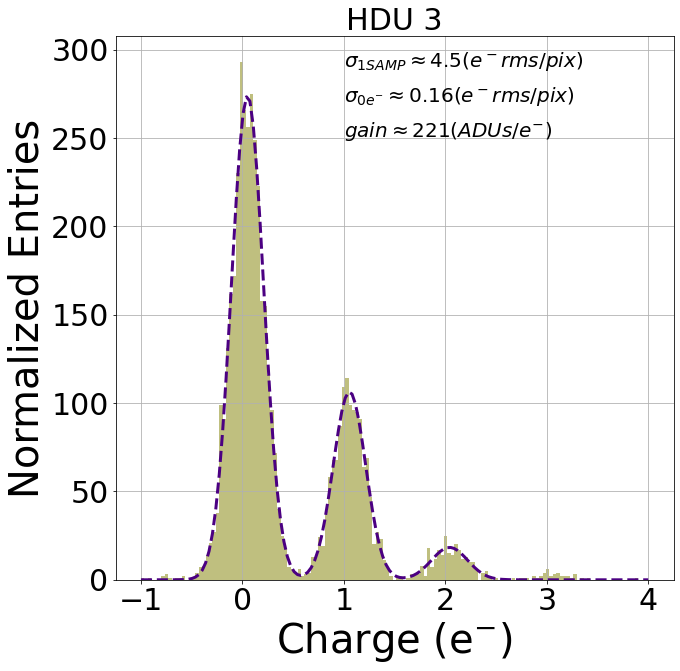}
    \caption{Readout noise performance of one of the packaged AstroSkipper CCDs. Left: Readout noise as a function of the number of samples for four extensions (HDUs) of one AstroSkipper CCD. There is a deviation from the expected theoretical expectation ($1/\sqrt{N_{samp}}$) in HDUs 1 and 3 starting at 300 samples per pixel. Right: Histogram of pixel values for HDU 3 (extension with the lowest readout noise) calculated from the average of 800 measurements per pixel (green histogram).
Single-electron resolution is achieved with a readout noise of $\sigma_{800} \sim 0.16 \ermspix$. 
We fit the peaks in the distribution with a multi-Gaussian model.
The gain of the CCD can be directly measured from the spacing between peaks.}
    \label{fig:noise}
\end{figure}

{\setlength{\parindent}{0cm}\textit{Readout noise:} We measure the readout noise performance of the packaged thick AstroSkipper detectors. Similar to  Drlica-Wagner, et al.\ (2020) \cite{10.1117/12.2562403}, we see a small light leak that leads to charge collection in the overscan pixels when they are shifting into the serial register, causing the appearance of electron peaks in the overscan pixel distribution. To measure the single-sample readout noise, we use a 800-sample image, where we read out 50 rows by 3200 columns of the detector, and fit the overscan pixel distribution with a multi-Gaussian model where the single-sample readout noise is given by the standard deviation of the 0\e peak. We perform the single-sample readout measurement for all of the amplifiers in the three detectors and find readout noise values ranging from $\sigma_{1} = 4.5 \ermspix$ to $\sigma_{1} = 10 \ermspix$. These initial readout noise values are higher than observed in similar Skipper CCDs, which have a single-sample readout noise of $\sim 3.5 \ermspix$ \cite{10.1117/12.2562403,Tiffenberg:2017, Rodrigues:2020}. The total readout noise in a system depends on all noise sources in the readout chain: the internal CCD readout MOS-FET transistor, interface electronics for preamplification, and data acquisition and video signal processing. Control and bias signals that can couple to the detector's internal readout and cabling, i.e., flex cables may also contribute to the readout noise. Furthermore external noise source such as vibrations from the cryogenic cooler and or poor grounding will couple to the video signals \cite{Cancelo:2020}. We are using newly assembled testing chambers and currently exploring these potential noise sources to improve the noise of the testing systems. The resulting noise performance as a function of the averaged samples is shown in the left panel of Fig.~\ref{fig:noise}. We see a deviation from the theoretical expectation (Eq. \ref{eqn:noise}) for extensions 1, and 3 at 300 samples per pixel (extensions with best readout noise performance). Extensions 0 and 2, which have poor noise performance, deviate from the theoretical expectation for all samples per pixel. We attribute these deviations to some source of correlated noise, which causes the readout noise behavior to deviate from the  theoretical model for uncorrelated measurements.
}

\noindent \textit{Electron counting and gain:}
The right plot in Fig.~\ref{fig:skipper} shows the pixel distribution of a 800-sample image taken with one of the the AstroSkipper CCDs; the peaks in the distribution quantize single electrons. Single electron/photon resolution in the detector is achieved when the charge of each pixel can be  precisely determined, i.e., electron-counting in Fig.~\ref{fig:noise} right panel. The gain of the entire signal video chain can be measured by the electron counting capability. The gain, which represents the relation between Analog-to-Digital-Units (ADUs) and electrons, is given by the difference of the mean values of consecutive fitted electron peaks in the pixel distribution. We measure gain values between 200 ADU/\e and 221 ADU/\e for all of the amplifiers in the three Skipper CCDs. 

\noindent \textit{Charge Transfer Inefficiency:} We measure the charge transfer inefficiency from two packaged AstroSkipper CCDs in all amplifiers, using the extended pixel edge response (EPER) \cite{Janesick:2001}. We use several single-sample images, containing $\sim 500 \e$; The EPER method consists of measuring the excess counts in the first overscan column resulting from CTI. This number is normalized to the total number of electrons in the last physical column and the number of horizontal charge shifts to get the final CTI value. We calculate an average CTI value, from all the amplifiers, of $5.3 \times 10^{-6}$. We note that this measurement did not consider voltage optimizations; however, CTI is one parameter we will optimize, along with full-well capacity and readout noise.    

\section{DEWAR AND FOCAL PLANE}
\label{sec:mechanics}
\subsection{Standard SOAR Dewar}
SOAR uses a standard dewar design for most of its visible instruments, which is also shared by some instruments at Cerro Tololo Inter-American Observatory (CTIO). The dewar has a single  LN2 tank, and it reaches a typical vacuum on the order of $10^{-6}$ Torr, with a holding time over 30 hours when full \cite{10.1117/12.457977}. It has a molecular sieve that allows it to keep a stable vacuum for several months without the need for intervention. 

\begin{figure}[t!]
    \centering
    \includegraphics[width=0.8\textwidth]{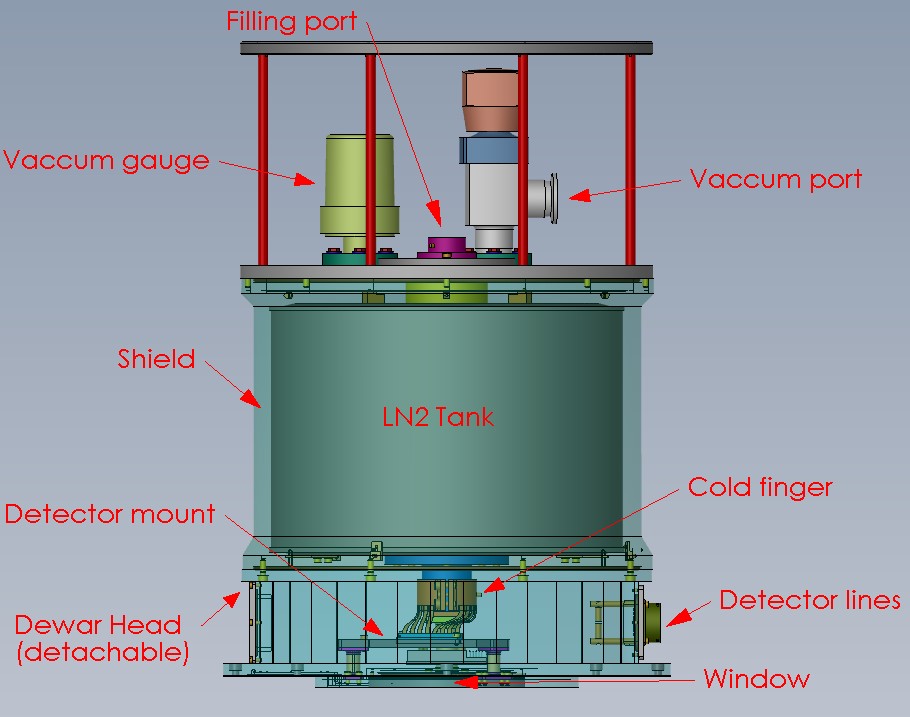}
    \caption{Cross section of a standard SOAR dewar. It uses LN2 to cool the detectors. The detector mount is attached to the tank through a cold finger and a copper piece, with braids trimmed for thermal adjustment.}
    \label{fig:dewar_cut}
\end{figure}

Fig. \ref{fig:dewar_cut} shows a cross section through a SOAR dewar. The LN2 tank is filled through the filling port once per day by the day crew. The dewar has an incorporated vacuum gauge that allows for both a direct view (digital screen) or remote telemetry of the current vacuum value. Surrounding the LN2 tank, there is an aluminum shield. The detector mount, located in the lower part of the dewar (``head''), is thermally connected to the LN2 tank through a copper piece attached to a coldfinger. The thermal adjustment is done by trimming some copper braids, ensuring a safe ``cold storage'' for the detector. The detector mount usually incorporates a heater that allows the CCDs to be kept at a prescribed operating temperature. The detector biases and clocks, as well as the video signals are passed through two 41-pin round hermetic connectors. The detector is installed in the lower part of the dewar, called the ``head'', which is detachable from the main body and can be customized depending on the instrument (back focal distance, detector mount, etc.).

\subsection{SIFS AstroSkipper Dewar}
\begin{figure}[t!]
    \centering
    \includegraphics[width=0.6\textwidth]{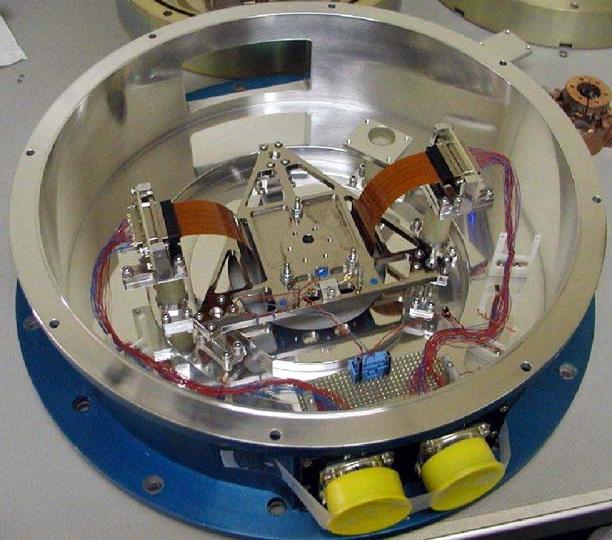}
    \vspace{1em}
    \caption{Current SIFS Detector mount, designed for a single e2v CCD44-82 ( 4 K$\times$4 K). The image show the detector, and below it, the mount, which is a triangular aluminum plate that attaches to the dewar through three posts. }
    \label{fig:current_e2v_mount}
\end{figure}

The SIFS dewar has a customized mount designed specifically for its e2v detector. Fig. \ref{fig:current_e2v_mount} shows the SIFS dewar head with the current detector  mount. The mount, which is the triangular aluminum piece in Fig.~\ref{fig:current_e2v_mount}, mounts to the dewar head through three posts.

Since in SIFS the dewar is independent of the rest of the instrument, by replacing the current detector and mount with one that holds the AstroSkipper CCDs and ensuring the same spatial coverage and back focal distance, there is no need to make any  modification to the instrument itself. However, given that SIFS is a production instrument and this is just a prototype, it is impossible to modify the SIFS dewar itself. Instead, we use an unfinished SOAR dewar with original parts built for the COSMOS red side (but never actually implemented), manufacture the missing pieces, and fit a spare dewar head identical to that of SIFS (this was kept in the detector laboratory for testing purposes). In this way, it will  be possible to remove the current SIFS dewar and install the one with the AstroSkipper CCDs without making any mechanical change to the instrument. This will also facilitate returning to the original dewar configuration, since this operation does not take more than a couple of hours during the day. 

\begin{figure}[t!]
    \centering
    \includegraphics[width=0.4\textwidth]{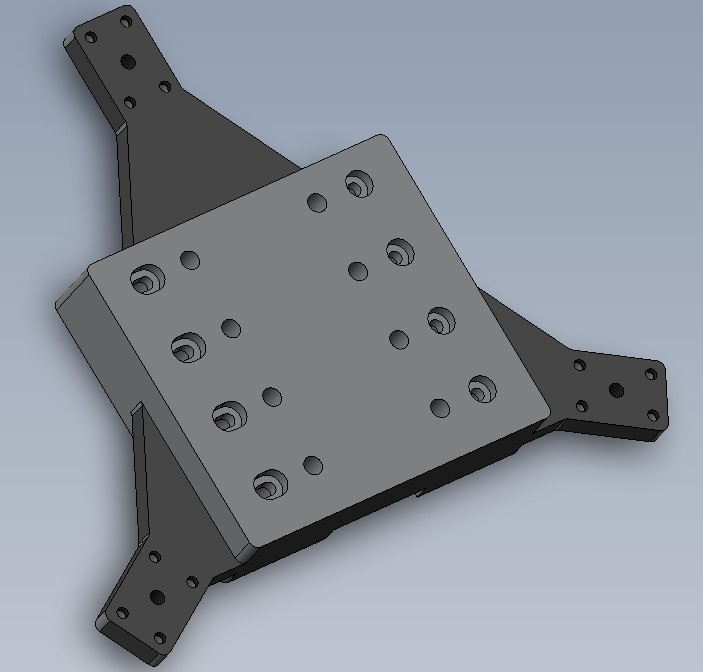}
    \vspace{1em}
    \caption{Designed mount for Skippe detectors. It is able to hold 4 AstroSkipper CCDs using the package designed at Fermilab. It uses three ears that allow to use the SIFS head mounting holes.}
    \label{fig:skippers_mount}
\end{figure}

Having a dewar and a head identical to that of SIFS, all that is required is to design and fabricate a new detector mount to hold the AstroSkipper CCDs. The mount was designed at LNA and manufactured at CTIO. The CAD model is shown in Fig.~\ref{fig:skippers_mount}. The mount is made of aluminum 6061 and holds four Fermilab-packaged AstroSkipper detectors using rods to place the detectors on the mount. The expected gap between detectors is expected to be $< 500 \um$, and any focal adjustment will be done with shims (the mount has no movable parts). The mount has three ears that are identical to those of the current mount, allowing it to use the same mounting holes. Special posts will be manufactured so that the mount is placed at a height that ensures the same back focal distance than the current detector. We will install two heaters and an RTD to have an appropriate temperature control. All the detector signals will be passed through the two 41-pin connectors.

\section{READOUT ELECTRONICS}
\label{sec:electronics}
\subsection{The LTA and Synchronized Readout Strategy}
The LTA readout board was designed at Fermilab for performing the readout of the Skipper CCDs \citep{Cancelo:2020}. We will be using 4 LTAs to read out the 4 Skipper CCDs that will go into the SIFS Skipper CCD focal plane. The LTA is a single PC board hosting 4 video channels for readout, plus CCD bias and control. The board was designed and optimized to work with $p$-channel, thick, high resistivity Skipper CCDs. The LTA is controlled by a Xilinx Artix XC7A200T FPGA, which sets up the programmable bias voltages/clocks, to move the charge along the CCD pixel array, video acquisition, telemetry, and data transfer from the board to the PC. The LTA software is written in C$++$ and allows for communication using UPD over IP through the computer giga-Ethernet port. The user can communicate with the LTA via terminal commands to perform board configuration, readout and telemetry requests, and sequencer uploading. The data acquisition comes in the form of images in FITS and other formats. 

\begin{figure}[t!]
    \centering
    \includegraphics[width=0.47\textwidth]{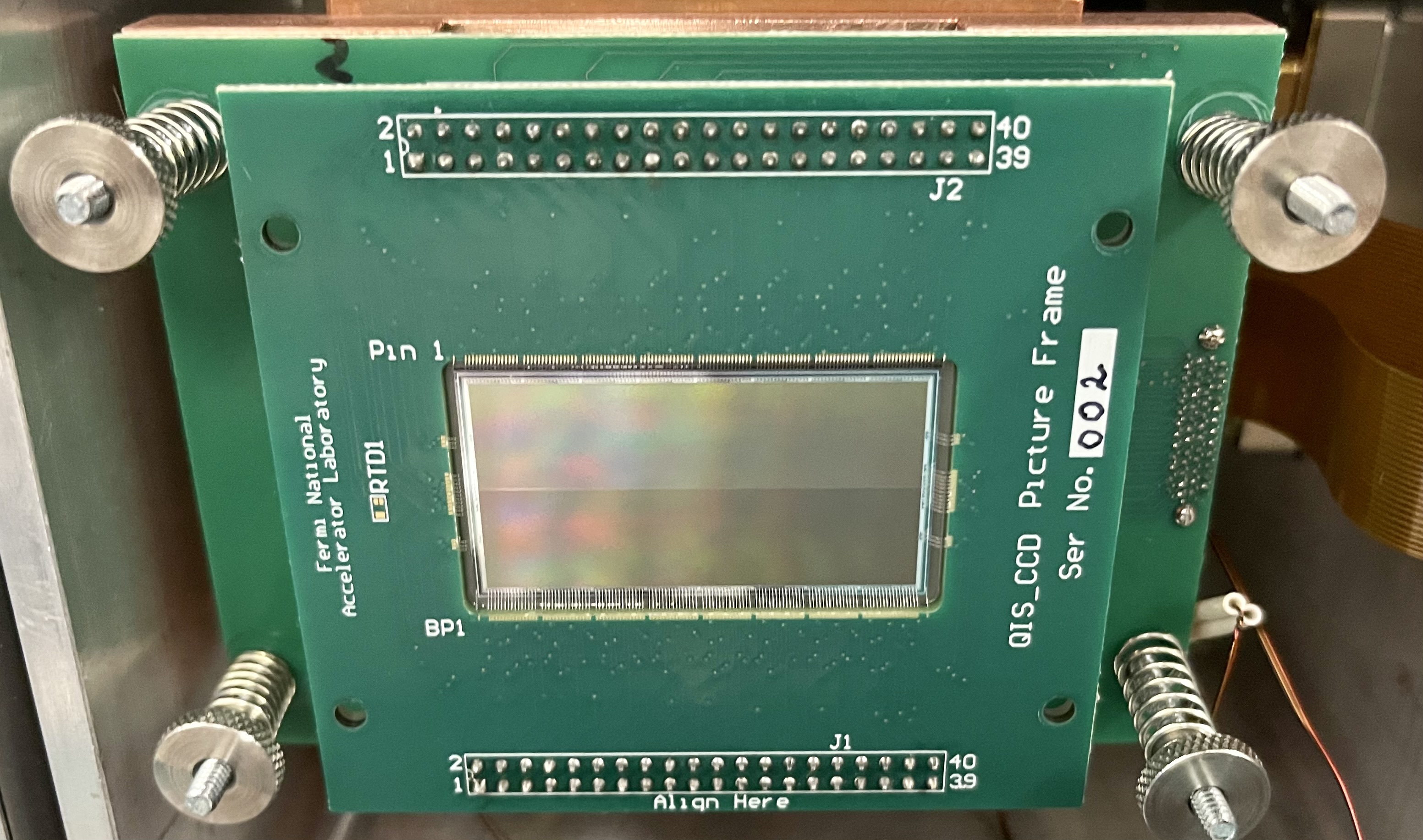}
    \includegraphics[width=0.32\textwidth]{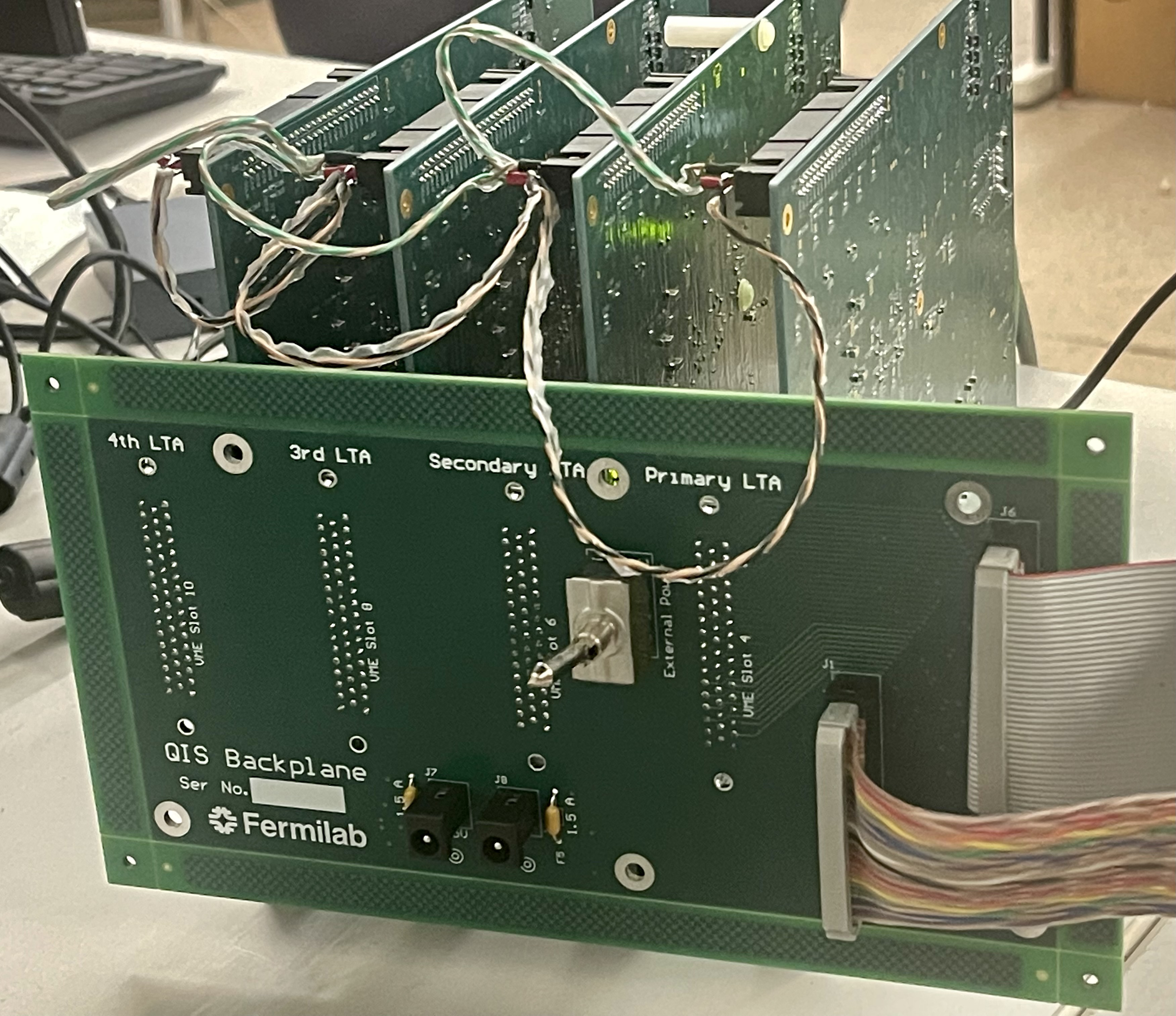}
    \vspace{1em}
    \caption{Left: 2K $\times$ 4K, 16 amplifier Skipper CCD. Amplifiers are placed along the long sides of the detector in groups of 8 amplifiers. Right: LTA synchronized readout system used for CCD testing. LTAs are docked to the backplane, which receives the video signals (rainbow ribbon cable) and biases and clocks (grey ribbon cable). Hardware synchronization between boards is done through the cables on the top of the boards.}
    \label{fig:skipper}
\end{figure}

The Skipper CCD focal plane will consists of four AstroSkipper CCDs; therefore, a synchronized readout system will be needed to read out the four detectors at the same time. The LTA allows for easily scaling to several CCDs in parallel. A synchronized LTA system consists of connected boards where the ``Leader'' LTA provides the voltage clocks and a start signal to the LTAs in ``Follower'' mode; this configuration assures that all boards commence the readout sequence in parallel with synchronized clocks. Each board will be responsible for reading out one detector (4 amplifiers per detector;  16 amplifiers total). We have extensively tested the LTA synchronized readout system, consisting of four LTAs, with a set of 16-amplifier, 2k $\times$ 4k detectors (Fig.~\ref{fig:skipper} Left), fabricated on the same wafers as the AstroSkipper CCD. For this synchronized system (Fig.~\ref{fig:skipper} Right), each of the four LTAs reads out a quadrant of the detector, containing four amplifiers; the LTAs connect to a backplane, which can provide power to the boards and distributes biases and video signals. 

\subsection{Readout Electronics at SOAR}
\subsubsection{LTA mechanical enclosure}\label{subsec:LTA enclosure}
\begin{figure}[t!]
    \centering
    \includegraphics[width=0.6\textwidth]{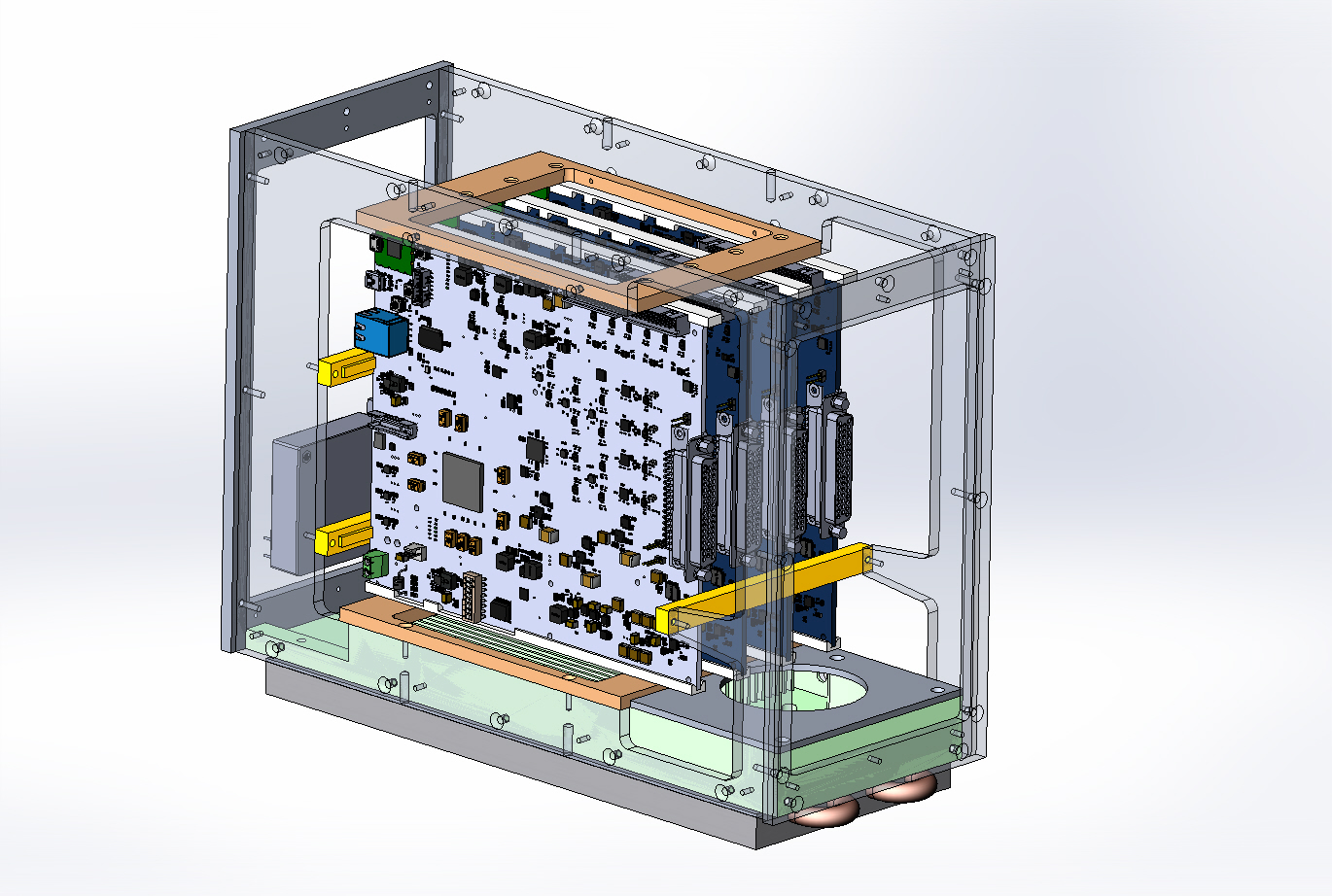}
    \caption{LTA enclosure box for SOAR. It includes a fan for air circulation and a heat exchanger plate, cooled with glycol going through the upper plate (shown at the bottom in the image). The box includes 12V and 5V supplies.}
    \label{fig:SOAR_lta_box}
\end{figure}

The current SIFS readout electronics is a SDSU-III controller. Given the thermal constrains, the housing is being cooled using glycol provided by the telescope facilities. The AstroSkipper CCDs, on the other hand, will be read out using 4 synchronized LTA boards. Due to the same environmental constrain, they will need to be enclosed in a glycol-cooled box. This box was designed at CTIO and is currently being  manufactured. A CAD model of the box can be seen in Fig.~\ref{fig:SOAR_lta_box}. The box has a fan for air re-circulation and a heat exchanger plate cooled with glycol which was designed to dissipate at least 200W. The box also includes two 12V supplies (6 amps each) and a 5V supply, used for the preamps on the electronic box inside the dewar.  

\subsubsection{Electronic preamplifier board and cabling}

Due to the nature of the application and the space available inside the dewar, CTIO designed a special preamplifier board that will both  fit on the envelope of the dewar and provide preamplification that is more suitable for an astronomical application.

\begin{figure}[t!]
    \centering
    \includegraphics[width=0.65\textwidth]{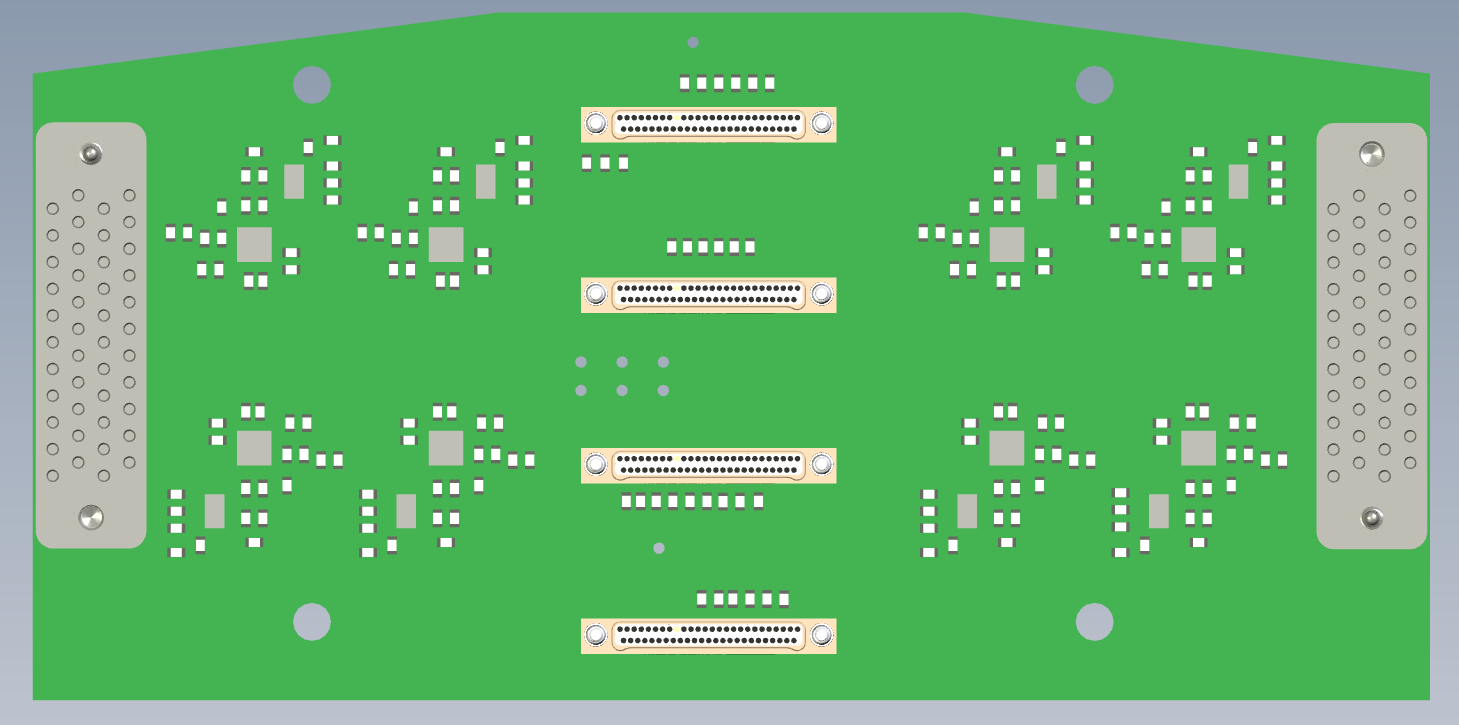}
    \vspace{1em}
    \caption{Top side CAD model of the preamplifier PCB board designed at CTIO showing eight of sixteen preamplifiers. The input connectors from the AstroSkipper CCDs are at the center and the output connectors are at the edges.}
    \label{fig:preamp_pcb}
\end{figure}

The electronic preamplifier PCB consists of a 12.49 cm $\times$ 6.15 cm, eight layer, two-sided PCB board with 16 preamplifier circuits, i.e., 4 preamplifiers per AstroSkipper CCD. The preamplifier board amplifies and converts to differential the output signal from the LSJ689 JFET, which allows it to isolate the video output signal from external noise sources. Finally, the differential output signals from the preamplifier board are connected to the input amplifiers at the LTA. A CAD model of the preamplifier board can be seen in Fig.~\ref{fig:preamp_pcb}. Currently the preamplifier PCB board is in the component assembly and testing stage.

For a nominal dynamic range of 50000 \e (more suitable for astronomical applications), the end-to-end gain from the detector to the ADC is set to 32.51 \gain. This gain give us a conversion factor of 0.48 \eadu in the ADC. Table \ref{tab:preamp} shows the simulated characteristics of the whole electronic chain from the CCD video output to the ADC input at the LTA. 

\begin{table}[t]
\centering
\caption{\label{tab:preamp}
Simulated electronic chain Parameters from the AstroSkipper CCD Video Output to the LTA Analog to Digital Converter (ADC).
}
\begin{tabular}{l c c c }
\hline
Characteristic  & Value  & Unit\\
\hline \hline
Passband Gain & 30.24 & \db \\
Low Cut-off & 15.9 & \hz  \\
High Cut-off & 960 & \khz  \\
Total Input Referred Noise
 & 2.018 & \erms\\
\hline
\end{tabular}
\end{table}

\begin{figure}[t!]
    \centering
    \includegraphics[width=0.65\textwidth]{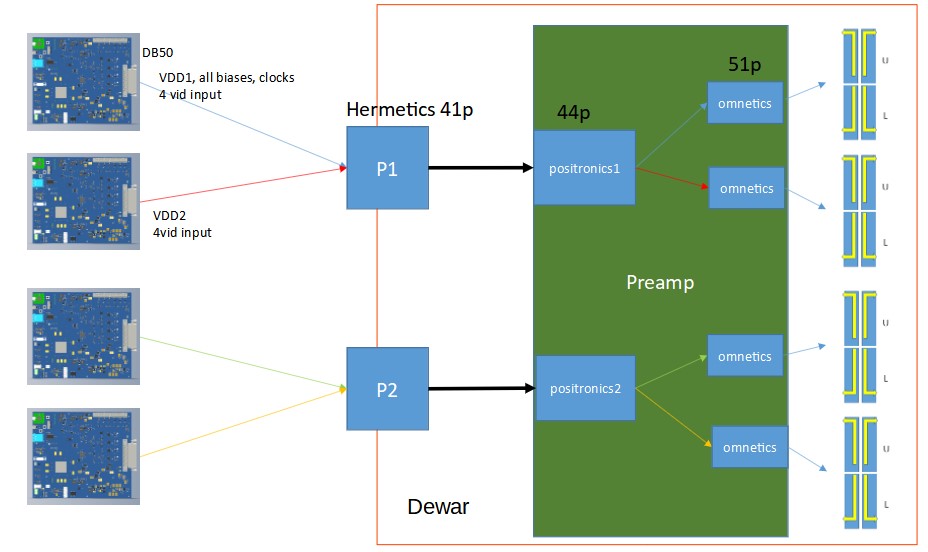}
    \caption{The wiring is designed so that two LTAs handle two AstroSkipper CCDs. This allows for symmetrical cables and independent working, if required for testing purposes. Biases and clocks are shared among two devices, except to VDD.}
    \label{fig:SOAR_skipper_wiring}
\end{figure}

The signals in and out of the preamplifier board are handled through cable rather than a VIB. This was for simplicity and compatibility with the SOAR dewar since the 2$\times$41 pin hermetics were enough to handle the signals for the 4 AstroSkipper detectors. There are two sets of cables, the internal, cryogenic ones, that take the signals from/to the preamplifier board to the hermetic connectors (inside the dewar), and another external set that takes them from the hermetics to the LTA boards inside the enclosure. The internal cables are made with 32-AWG cryogenic phosphor-bronze Lakeshore cable, due to its good combination of low resistivity (3.45 $\Omega$/m) and low thermal conductivity (25 W/mK) at 77K \cite{LakeshoreWebSite}, using single strand for clocks and biases, and dual twisted pair ones for video signals. The external cables are single 32-AWG copper cables for biases and clocks, and 32-AWG twisted pairs for video signals. Fig.~\ref{fig:SOAR_skipper_wiring} shows the wiring design. Two LTAs handle two AstroSkipper detectors independently, with one hermetic servicing two AstroSkipper CCDs and its LTAs. This allows for symmetrical cables and also, if required, it would be possible to read any group of only two AstroSkipper CCDs. The biases and clocks are shared between two AstroSkipper CCDs and are provided by one LTA, except for VDD which is independent. The four LTAs will be installed on the electronics box described in Section~\ref{subsec:LTA enclosure}.

\section{Future Work}
\label{sec:outlook}
\subsection{Detector Characterization}
\label{sec:optical}
LBNL is currently processing five wafers, which includes thinning to 250 $\mu$m, backside treatment, and the application of antireflective coating \cite{Flaugher:2015}.
We will perform extensive optical characterization of these processed AstroSkipper CCDs. We plan to use an optical system, which includes a tungsten arc lamp, motorized filter wheel, monochromator, shutter, NIST-calibrated photodiode and integrating sphere. This system has been assembled and tested; we have placed the optical setup in a ``dark room'' to minimize the light leaks and background ambient radiation. We plan to study the full dynamic range of these detectors by constructing photon transfer curves (PTCs); PTCs will also serve as a tool to optimize the full well capacity. Furthermore, we plan to perform an absolute quantum efficiency (QE) measurement by packaging a photodiode and placing it inside the vacuum testing chamber for calibration; we have been able to do relative QE measurements on Skipper CCDs \cite{10.1117/12.2562403}. These studies will demonstrate the feasibility for using the Skipper CCD at SIFS as we expect them to retain the well-characterized performance of conventional CCDs in the optical, near infrared regimes.       

\subsection{Fast Readout Strategies and Optimization}
The Skipper CCD's ability to reduce readout noise through  multiple non-destructive measurements of charge packets comes at the cost of increased readout times. For instance, Skipper CCD readout times for a single sample measurement with readout noise of $\sigma_1 \sim 3.5$ \ermspix is $\sim 40 \mu s /$pix. Yet, when taking multiple non-destructive measurements of the charge in each pixel, the readout time scales as $t \propto N_{\rm samp} \propto 1/\sigma_{N}^{2}$. Therefore, in applications where ultra-low noise is required, data collection can take several hours \cite{Barak:2020}. Long readout times are undesirable in astronomical observations. For nearly all astronomical applications, the exposure time and the readout time need to be optimized as a function of the signal-to-noise. Therefore, for a given astronomical observation, there is a minimum at which the signal-to-noise gain from the Skipper CCD's reduced readout noise is overcome by the lost exposure time due to a long readout time \cite{10.1117/12.2562403}. We will explore software and firmware based strategies that can be implemented in the time frame of the installation of Skipper CCD focal plane at SIFS.

{\setlength{\parindent}{0cm} \textit{Firmware Modifications:}
The LTA's FPGA firmware enables complex clock sequences; however for fast sequences, this flexibility can result in broken timing structures for time-critical clocking signals and data corruption \cite{Chierchie_2021}. The current readout rate for simple clock sequences such as the multiple samples per pixel readout mode is limited to $\sim 200$ kpix/sec. We have developed a new low-level LTA firmware that allows for sequence executions at a rate of $> 1$ Mpix/sec; there is extensive testing that needs to be completed in order to validate the detector readout noise performance running with this new firmware. Furthermore, we need to understand and remove any systematics, such as bias shifts, introduced to the data as a result of reading out fast.     

\textit{Region of Interest:} The Skipper CCD provides the ability to adjust the number of non-destructive measurements of charge packets on a pixel-by-pixel basis. Functionally, one can configure regions of interest (ROI) on the detector to readout out with ultra-low noise while the remaining of the detector is read out quickly with a single sample per pixel \cite{10.1117/12.2562403}. Furthermore, the number of non-destructive measurements can be chosen dynamically depending on the measured charge contained in a pixel (also called ``energy of interest'' in the context of energetic particle detection) \cite{Chierchie_2021}. The ROI technique can drastically reduced readout times since depending on the application, one might need to read out a portion of the detector with low noise. For spectroscopy, one can define a ROI to be a faint spectral line at a predetermined wavelength and read out this region with a target samples per pixel to achieve the desired noise. We plan to make this functionality more robust by designing software to target pre-determined regions of the CCD area suited for spectroscopy. Furthermore, we need to develop correction techniques to bias-correct images where the readout noise varies pixel-to-pixel. We will implement ROI at SIFS for the first time in a real observing scenario as a proof of concept.      

}

\subsection{Detector Testing at CTIO  and SOAR}

We will conduct extensive characterization of the performance of the four packaged, 250$\mu m$ thick AstroSkipper detectors at the CTIO detector's lab to verify and compare with the results from Fermilab characterization tests. This will allow us to test for any damage of the AstroSkipper detectors during transport to Chile; furthermore, it is important to verify previous characterization results obtained at Fermilab before installing the AstroSkipper CCD focal plane at SOAR.

At CTIO detector's lab, we will use the implemented  fully-automated detector testbench, an evolution of the testbench constructed in collaboration between Fermilab and CTIO for testing DESI detectors. This testbench, described in Bonati et al.\ (2020) \cite{10.1117/12.2559203}, is used to perform detector characterization tests at different illumination levels. The testbench uses an automated system consisting of optical hardware similar to the one we use at Fermilab for optical characterization (Section \ref{sec:optical}) and software that allows for data acquisition, reduction, and visualization. The automated testbench characterization sequence is divided between light and dark tests. Light tests measure low- and high-light non linearity, to evaluate the non-linearity of the detector in low-, high-light conditions, Charge Transfer Efficiency (CTE), flat-fields, for identifying cosmetic defects, and absolute quantum efficiency, to measure how many photons that arrived at the detector are effectively converted to electrons. Dark tests, performed with no light, measure the readout noise of the detector, and dark current. Test results are compared with predetermined requirements to assign a grade, i.e., mechanical, engineering, and science to the detector; a science grade-detector must have QE $>0.80$, cosmetics $<0.1$, CTI $<10^{-5}$, full-well $> 130$ K $\e$, non-linearity $<1.5$, dark current $<30 \e$/pix/h, and readout noise $<15\e$ at 100kHz \cite{10.1117/12.2559203}. CTIO detector's lab has a similar testing chamber to what we have at Fermilab; therefore, testing individual detectors should be possible. However, we plan to assemble the AstroSkipper CCD focal plane in the dewar and test the four synchronized AstroSkipper CCDs at CTIO detector's lab. We will then move the working assembly and detector configuration to the SOAR telescope. 

Detector characterization at SOAR will focus on validating target single-sample readout noise and readout time for observation optimization purposes. For example, we will verify that the 16-amplifiers (4 amplifiers per AstroSkipper CCD) achieve a single-sample readout noise $\leq 3.5$ \ermspix with readout time of $20 \mu$s per pixel ($\sim$ 30s readout time for the entire detector array of four AstroSkipper CCDs). We will also target a readout noise of $<1$\ermspix in $< 10$min through multiple measurements of the charge in each pixel. Furthermore, we will verify the ROI capability by demonstrating that the AstroSkipper CCDs can achieve sub-electron readout noise over a fraction of the detectors' area.  

Note that the DAQ software that will be used at CTIO's laboratory testbench, called panView\cite{10.1117/12.461433}\cite{SDFA}, allows to hide the particulars of the detector controller, making it transparent for the callers what hardware is in use. Since the SIFS instrument at the SOAR telescope uses panView as DAQ (using an SDSU-III controller), there are no modifications of any type that will be required to the instrument software (GUI/scripts/TCS/mechanisms, etc.).

\subsection{SIFS Commissioning and Installation }
Commission of the Skipper CCD focal plane prototype will consists of removing the current cryostat from the bench spectrograph (Fig.~\ref{fig:spectrograph}), installing the new cryostat with the four mounted AstroSkipper CCDs, proper alignment to ensure focal distance, and installation and configuration of the readout electronics. Furthermore, we will configure detector operation modes that are optimized for observing scenarios. Detector optimization will involve developing tools to calculate and optimize the trade-off between readout noise and readout time as a function of observed wavelength, signal strength, and background level in order to determine the optimal number of non-destructive measurements per pixel. Moreover, as part of the observation configurations, we will test ROI selection for faster readout times in observational scenarios where faint spectral features are targeted at predetermined pixel locations. We will characterize detector performance in operation at the telescope; this will involve exposing the detector to different signal strengths, i.e., bright and faint spectral lines and quantify readout noise and detector cross-talk, for example. Finally, we will demonstrate stable detector performance over extended observing runs.

\subsection{Planned Science Verification}
In science verification, we intend to validate the performance of the Skipper CCD by targeting faint sources at blue wavelengths where the sky background is minimal and the detector readout noise contribution is important. For example, one plausible science verification target would be the study of emission lines in a population of star-forming low-surface brightness galaxies (LSBGs) detected in multi-band imaging from the Dark Energy Survey \cite{Tanoglidis:2021}. We will follow the procedure described in Greco et al.\ (2018) \cite{Greco:2018b} to select likely star-forming LSBGs starting with the blue subpopulation and matching to UV detections in {\it GALEX}. We will then use the procedure described in Palumbo et al.\ (2020) \cite{Palumbo:2020} to extract object's physical properties from SIFS data cubes after having selected target LSBGs. Our goal is to demonstrate that the AstroSkipper will reduce readout noise in the detector pixels corresponding to the expected spatial and spectral location of star-forming H\;{\sc ii} regions in the targeted LSBGs. We expect to be able to recover redshifts, internal velocity dispersions, and star formation rates from the data taken with the AstroSkipper CCD focal plane at SIFS.


\acknowledgments 
 
This work was supported in part by Fermilab LDRD 2019.011 and LDRD 2022.053. 
This material is based upon work supported by the U.S. Department of Energy, Office of Science, Office of Workforce Development for Teachers and Scientists, Office of Science Graduate Student Research (SCGSR) program. The SCGSR program is administered by the Oak Ridge Institute for Science and Education for the DOE under contract number DE‐SC0014664. 
The work of M.B., B.C., and P.M. is supported by NOIRLab, which is managed by the Association of Universities for Research in Astronomy (AURA) under a cooperative agreement with the National Science Foundation. 
 SIFS construction was supported by FAPESP 1999/03744-1 and Laboratório Nacional de Astrofísica (LNA/MCTI).
CCD development was supported by the Lawrence Berkeley National Laboratory Director, Office of Science, of the U.S.\ Department of Energy under Contract No.\ DE-AC02-05C
H11231. 
This manuscript has been authored by Fermi Research Alliance, LLC under Contract No.\ DE-AC02-07CH11359 with the U.S.\ Department of Energy, Office of Science, Office of High Energy Physics. 
The United States Government retains and the publisher, by accepting the article for publication, acknowledges that the United States Government retains a non-exclusive, paid-up, irrevocable, world-wide license to publish or reproduce the published form of this manuscript, or allow others to do so, for United States Government purposes.

\renewcommand{\refname}{REFERENCES}

\bibliography{main} 

\begin{thebibliography}{10}

\bibitem{10.1117/12.857698}
de~Oliveira, A.~C., de~Oliveira, L.~S., Gneiding, C.~D., Barbuy, B., Jones, D.,
  Figueredo, M.~V., Lépine, J. R.~D., Macanhan, V. B.~P., de~Oliveira, J.
  B.~C., and Taylor, K., ``{The SOAR integral field unit spectrograph optical
  design and IFU implementation},'' in [{\em Modern Technologies in Space- and
  Ground-based Telescopes and Instrumentation}{\nolinebreak\hspace{0.1em}]},
  Atad-Ettedgui, E. and Lemke, D., eds.,  {\bf 7739},  1646 -- 1657,
  International Society for Optics and Photonics, SPIE (2010).

\bibitem{da_Silva_2020}
da~Silva, P., Menezes, R.~B., and Steiner, J.~E., ``The nuclear region of {NGC}
  613 {\textendash} i. multiwavelength analysis,'' {\em Monthly Notices of the
  Royal Astronomical Society}~{\bf 492},  5121--5140 (jan 2020).

\bibitem{10.1117/12.461977}
Lepine, J. R.~D., de~Oliveira, A.~C., Figueredo, M.~V., Castilho, B.~V.,
  Gneiding, C., Barbuy, B., Jones, D.~J., Kanaan, A., de~Oliveira, C.~M.,
  Strauss, C., Rodrigues, F., Andrade, C.~R., de~Oliveira, L.~S., and
  de~Oliveira, J.~B., ``{SIFUS: SOAR integral field unit spectrograph},'' in
  [{\em Instrument Design and Performance for Optical/Infrared Ground-based
  Telescopes}{\nolinebreak\hspace{0.1em}]},  Iye, M. and Moorwood, A. F.~M.,
  eds.,  {\bf 4841},  1086 -- 1095, International Society for Optics and
  Photonics, SPIE (2003).

\bibitem{Janesick:1990}
{Janesick}, J.~R., {Elliott}, T., {Dingizian}, A., {Bredthauer}, R.~A., and
  {Chandler}, C.~E., ``{New advancements in charge-coupled device technology -
  Sub-electron noise and 4096 $\times$ 4096 pixel CCDs},'' {\em SPIE}~{\bf
  1242},  223--237 (1990).

\bibitem{10.1117/12.19457}
Chandler, C.~E., Bredthauer, R.~A., Janesick, J.~R., and Westphal, J.~A.,
  ``{Sub-electron noise charge-coupled devices},'' in [{\em Charge-Coupled
  Devices and Solid State Optical Sensors}{\nolinebreak\hspace{0.1em}]},
  Blouke, M.~M., ed.,  {\bf 1242},  238 -- 251, International Society for
  Optics and Photonics, SPIE (1990).

\bibitem{Tiffenberg:2017}
Tiffenberg, J., Sofo-Haro, M., Drlica-Wagner, A., Essig, R., Guardincerri, Y.,
  Holland, S., Volansky, T., and Yu, T.-T., ``{Single-electron and
  single-photon sensitivity with a silicon Skipper CCD},'' {\em Phys. Rev.
  Lett.}~{\bf 119}(13),  131802 (2017).

\bibitem{Cancelo:2020}
{Cancelo}, G., {Chavez}, C., {Chierchie}, F., {Estrada}, J., {Fernandez
  Moroni}, G., {Paolini}, E.~E., {Sofo Haro}, M., {Soto}, A., {Stefanazzi}, L.,
  {Tiffenberg}, J., {Treptow}, K., {Wilcer}, N., and {Zmuda}, T., ``{Low
  Threshold Acquisition controller for Skipper CCDs},'' {\em IEEE 2019
  Argentine Conference on Electronics (CAE)} ,  86--91 (2020).

\bibitem{10.1117/1.JATIS.7.1.015001}
Cancelo, G.~I., Chavez, C., Chierchie, F., Estrada, J., Fernandez-Moroni, G.,
  Paolini, E.~E., Haro, M.~S., Soto, A., Stefanazzi, L., Tiffenberg, J.,
  Treptow, K., Wilcer, N., and Zmuda, T.~J., ``{Low threshold acquisition
  controller for Skipper charge-coupled devices},'' {\em Journal of
  Astronomical Telescopes, Instruments, and Systems}~{\bf 7}(1),  1 -- 19
  (2021).

\bibitem{Barak_2022}
Barak, L., Bloch, I.~M., Botti, A., Cababie, M., Cancelo, G., Chaplinsky, L.,
  Chierchie, F., Crisler, M., Drlica-Wagner, A., Essig, R., Estrada, J.,
  Etzion, E., Moroni, G.~F., Gift, D., Holland, S.~E., Munagavalasa, S., Orly,
  A., Rodrigues, D., Singal, A., Haro, M.~S., Stefanazzi, L., Tiffenberg, J.,
  Uemura, S., Volansky, T., and and, T.-T.~Y., ``{SENSEI}: Characterization of
  single-electron events using a skipper charge-coupled device,'' {\em Physical
  Review Applied}~{\bf 17} (jan 2022).

\bibitem{10.1117/12.2562403}
Drlica-Wagner, A., Villalpando, E.~M., O'Neil, J., Estrada, J., Holland, S.,
  Kurinsky, N., Li, T., Moroni, G.~F., Tiffenberg, J., and Uemura, S.,
  ``{Characterization of skipper CCDs for cosmological applications},'' in
  [{\em X-Ray, Optical, and Infrared Detectors for Astronomy
  IX}{\nolinebreak\hspace{0.1em}]},  Holland, A.~D. and Beletic, J., eds.,
  {\bf 11454},  210 -- 223, International Society for Optics and Photonics,
  SPIE (2020).

\bibitem{Dawson:2008}
{Dawson}, K., {Bebek}, C., {Emes}, J., {Holland}, S., {Jelinsky}, S.,
  {Karcher}, A., {Kolbe}, W., {Palaio}, N., {Roe}, N., {Saha}, J., {Takasaki},
  K., and {Wang}, G., ``{Radiation Tolerance of Fully-Depleted P-Channel CCDs
  Designed for the SNAP Satellite},'' {\em IEEE Transactions on Nuclear
  Science}~{\bf 55},  1725--1735 (June 2008).

\bibitem{Flaugher:2015}
Flaugher, B. et~al., ``{The Dark Energy Camera},'' {\em Astron. J.}~{\bf 150},
  150 (2015).

\bibitem{10.1117/12.856593}
Macanhan, V. B.~P., Santoro, F.~G., Gneiding, C.~D., de~Oliveira, A.~C.,
  Lourenço, F., Barbuy, B., Lépine, J. R.~D., Figueiredo, M.~V., Silva,
  P.~F., Castilho, B., Ribeiro, F.~F., de~Arruda, M.~V., Gutierrez, A.~M.,
  Zambretti, L.~R., Rodrigues, F., Luz, H. D. P.~D., and da~Silva, J.~M.,
  ``{Mechanical design of SIFS SOAR integral field unit spectrograph},'' in
  [{\em Ground-based and Airborne Instrumentation for Astronomy
  III}{\nolinebreak\hspace{0.1em}]},  McLean, I.~S., Ramsay, S.~K., and Takami,
  H., eds.,  {\bf 7735},  2439 -- 2445, International Society for Optics and
  Photonics, SPIE (2010).

\bibitem{Palumbo:2020}
{Palumbo}, Michael~L., I., {Kannappan}, S.~J., {Frazer}, E.~M., {Eckert},
  K.~D., {Norman}, D.~J., {Fraga}, L., {Quint}, B.~C., {Amram}, P., {Mendes de
  Oliveira}, C., {Bittner}, A.~S., {Moffett}, A.~J., {Stark}, D.~V., {Norris},
  M.~A., {Cleaves}, N.~T., and {Carr}, D.~S., ``{Linking compact dwarf
  starburst galaxies in the RESOLVE survey to downsized blue nuggets},'' {\em
  \mnras}~{\bf 494},  4730--4750 (June 2020).

\bibitem{Rodrigues:2020}
{Rodrigues}, D., {Andersson}, K., {Cababie}, M., {Donadon}, A., {Botti}, A.,
  {Cancelo}, G., {Estrada}, J., {Fernandez-Moroni}, G., {Piegaia}, R.,
  {Senger}, M., {Sofo Haro}, M., {Stefanazzi}, L., {Tiffenberg}, J., and
  {Uemura}, S., ``{Absolute measurement of the Fano factor using a
  Skipper-CCD},'' {\em arXiv e-prints} ,  arXiv:2004.11499 (2020).

\bibitem{Derylo:2006}
{Derylo}, G., {Diehl}, H.~T., and {Estrada}, J., ``{0.250mm-thick CCD packaging
  for the Dark Energy Survey Camera array},'' in [{\em Society of Photo-Optical
  Instrumentation Engineers (SPIE) Conference
  Series}{\nolinebreak\hspace{0.1em}]},  {\em Society of Photo-Optical
  Instrumentation Engineers (SPIE) Conference Series} {\bf 6276},  627608
  (2006).

\bibitem{Janesick:2001}
{{Janesick}, J.~R.},  [{\em {Scientific Charge Coupled
  Devices}}{\nolinebreak\hspace{0.1em}]}, SPIE Publications (2001).

\bibitem{10.1117/12.457977}
Walker, A.~R., Boccas, M., Bonati, M., Galvez, R., Martinez, M., Schurter, P.,
  Schmidt, R.~E., Ashe, M.~C., Delgado, F., and Tighe, R., ``{SOAR Optical
  Imager},'' in [{\em Instrument Design and Performance for Optical/Infrared
  Ground-based Telescopes}{\nolinebreak\hspace{0.1em}]},  Iye, M. and Moorwood,
  A. F.~M., eds.,  {\bf 4841},  286 -- 294, International Society for Optics
  and Photonics, SPIE (2003).

\bibitem{LakeshoreWebSite}
Lakeshore, ``Cryogenic-wire.''
  \url{https://www.lakeshore.com/products/categories/specification/temperature-products/cryogenic-accessories/cryogenic-wire}.

\bibitem{Barak:2020}
{Barak}, L., {Bloch}, I.~M., {Cababie}, M., {Cancelo}, G., {Chaplinsky}, L.,
  {Chierchie}, F., {Crisler}, M., {Drlica-Wagner}, A., {Essig}, R., {Estrada},
  J., {Etzion}, E., {Fernandez Moroni}, G., {Gift}, D., {Munagavalasa}, S.,
  {Orly}, A., {Rodrigues}, D., {Singal}, A., {Sofo Haro}, M., {Stefanazzi}, L.,
  {Tiffenberg}, J., {Uemura}, S., {Volansky}, T., and {Yu}, T.-T., ``{SENSEI:
  Direct-Detection Results on sub-GeV Dark Matter from a New Skipper-CCD},''
  {\em Phys. Rev. Lett.}~{\bf 125},  171802 (2020).

\bibitem{Chierchie_2021}
Chierchie, F., Moroni, G.~F., Stefanazzi, L., Paolini, E., Tiffenberg, J.,
  Estrada, J., Cancelo, G., and Uemura, S., ``Smart readout of nondestructive
  image sensors with single photon-electron sensitivity,'' {\em Physical Review
  Letters}~{\bf 127} (dec 2021).

\bibitem{10.1117/12.2559203}
Bonati, M., Estrada, J., Castaneda, A., and Hernandez, P., ``{Fully automated
  detector testbench},'' in [{\em Software and Cyberinfrastructure for
  Astronomy VI}{\nolinebreak\hspace{0.1em}]},  Guzman, J.~C. and Ibsen, J.,
  eds.,  {\bf 11452},  290 -- 302, International Society for Optics and
  Photonics, SPIE (2020).

\bibitem{10.1117/12.461433}
Ashe, M.~C., Bonati, M., and Heathcote, S., ``{ArcVIEW: a LabVIEW-based
  astronomical instrument control system},'' in [{\em Advanced Telescope and
  Instrumentation Control Software II}{\nolinebreak\hspace{0.1em}]},  Lewis,
  H., ed.,  {\bf 4848},  508 -- 518, International Society for Optics and
  Photonics, SPIE (2002).

\bibitem{SDFA}
Astrophysics and space~science library, eds.,  [{\em Scientific Detectors for
  Astronomy, The beginning of a New Era}{\nolinebreak\hspace{0.1em}]},
  vol.~300, ch.~IV, ISPI's software - an application of the ArcVIEW system,
  427--430, Kluwer Academic Publishers (2005).

\bibitem{Tanoglidis:2021}
{Tanoglidis}, D., {Drlica-Wagner}, A., {Wei}, K., {Li}, T.~S., {S{\'a}nchez},
  J., {Zhang}, Y., {Peter}, A.~H.~G., {Feldmeier-Krause}, A., {Prat}, J.,
  {Casey}, K., {Palmese}, A., {S{\'a}nchez}, C., {DeRose}, J., {Conselice}, C.,
  {Gagnon}, L., {Abbott}, T.~M.~C., {Aguena}, M., {Allam}, S., {Avila}, S.,
  {Bechtol}, K., {Bertin}, E., {Bhargava}, S., {Brooks}, D., {Burke}, D.~L.,
  {Rosell}, A.~C., {Kind}, M.~C., {Carretero}, J., {Chang}, C., {Costanzi}, M.,
  {da Costa}, L.~N., {De Vicente}, J., {Desai}, S., {Diehl}, H.~T., {Doel}, P.,
  {Eifler}, T.~F., {Everett}, S., {Evrard}, A.~E., {Flaugher}, B., {Frieman},
  J., {Garc{\'\i}a-Bellido}, J., {Gerdes}, D.~W., {Gruendl}, R.~A., {Gschwend},
  J., {Gutierrez}, G., {Hartley}, W.~G., {Hollowood}, D.~L., {Huterer}, D.,
  {James}, D.~J., {Krause}, E., {Kuehn}, K., {Kuropatkin}, N., {Maia},
  M.~A.~G., {March}, M., {Marshall}, J.~L., {Menanteau}, F., {Miquel}, R.,
  {Ogando}, R.~L.~C., {Paz-Chinch{\'o}n}, F., {Romer}, A.~K., {Roodman}, A.,
  {Sanchez}, E., {Scarpine}, V., {Serrano}, S., {Sevilla-Noarbe}, I., {Smith},
  M., {Suchyta}, E., {Tarle}, G., {Thomas}, D., {Tucker}, D.~L., {Walker},
  A.~R., and {DES Collaboration}, ``{Shadows in the Dark:
  Low-surface-brightness Galaxies Discovered in the Dark Energy Survey},'' {\em
  \apjs}~{\bf 252},  18 (Feb. 2021).

\bibitem{Greco:2018b}
{Greco}, J.~P., {Goulding}, A.~D., {Greene}, J.~E., {Strauss}, M.~A., {Huang},
  S., {Kim}, J.~H., and {Komiyama}, Y., ``{A Study of Two Diffuse Dwarf
  Galaxies in the Field},'' {\em \apj}~{\bf 866},  112 (Oct. 2018).

\end{thebibliography}
\bibliographystyle{spiebib} 

\end{document}